\documentclass[usenatbib]{mn2e}

\usepackage{epsfig}
\usepackage{mathrsfs,aas_macros}

\def\Lya{Ly$\alpha$~} 
\def\Lyb{Ly$\beta$~}
\def\Lyg{Ly$\gamma$~}
\def\Lyd{Ly$\delta$~}

\def\HI{\hbox{H$\,\rm \scriptstyle I\ $}}

\def\HeII{\hbox{He$\,\rm \scriptstyle II\ $}}
\def\HeIII{\hbox{He$\,\rm \scriptstyle III\ $}} 
 
\def\CIII{\hbox{C$\,\rm \scriptstyle III\ $}}
\def\CIV{\hbox{C$\,\rm \scriptstyle IV\ $}}

\def\OVI{\hbox{O$\,\rm \scriptstyle VI\ $}} 
\def\SiIV{\hbox{Si$\,\rm \scriptstyle IV\ $}}

\title[UVB fluctuations: carbon and silicon]{The impact of spatial
  fluctuations in the ultra-violet background on intergalactic carbon
  and silicon}

\author[J.S. Bolton \& M. Viel] {James S. Bolton$^{1}$ \& Matteo
  Viel$^{2,3}$\\ $^1$ School of Physics, University of Melbourne,
  Parkville, VIC 3010, Australia \\ $^2$ INAF-Osservatorio Astronomico
  di Trieste, Via G. B. Tiepolo 11, I-34131 Trieste, Italy\\ $^3$
  INFN/National Institute for Nuclear Physics, Via Valerio 2, I-34127
  Trieste, Italy}

\begin{document}

\date{}

\maketitle

\label{firstpage}

\begin{abstract}

Spatial inhomogeneities in the spectral shape of the ultra-violet
background (UVB) at the tail-end of \HeII reionisation are thought to
be the primary cause of the large fluctuations observed in the \HeII
to \HI \Lya forest optical depth ratio, $\tau_{\rm HeII}/\tau_{\rm
  HI}$, at $z\simeq 2-3$.  These spectral hardness fluctuations will
also influence the ionisation balance of intergalactic metals; we
extract realistic quasar absorption spectra from a large
hydrodynamical simulation to examine their impact on intergalactic
\SiIV and \CIV absorbers.  Using a variety of toy UVB models, we find
that while the predicted spatial inhomogeneities in spectral hardness
have a significant impact on $\tau_{\rm HeII}/\tau_{\rm HI}$, the
longer mean free path for photons with frequencies above and below the
\HeII ionisation edge means these fluctuations have less effect on the
\SiIV and \CIV ionisation balance.  Furthermore, UVB models which
produce the largest fluctuations in specific intensity at the \HeII
ionisation edge also have the softest ionising spectra, and thus
result in photo-ionisation rates which are too low to produce
significant fluctuations in the observed $\tau_{\rm SiIV}/\tau_{\rm
  CIV}$.  Instead, we find spatial variations in the IGM metallicity
will dominate any scatter in $\tau_{\rm SiIV}/\tau_{\rm CIV}$.  Our
results suggest that observational evidence for homogeneity in the
observed $\tau_{\rm SiIV}/\tau_{\rm CIV}$ distribution does not rule
out the possibility of significant fluctuations in the UVB spectral
shape at $z\simeq 2-3$.  On the other hand, the scatter in metallicity
inferred from observations of intergalactic \CIV and \SiIV absorption
at $z\simeq 2-3$ using spatially uniform ionisation corrections is
likely intrinsic, and therefore provides a valuable constraint on
intergalactic metal enrichment scenarios at these redshifts.

\end{abstract}
 
\begin{keywords}
  methods: numerical - intergalactic medium - quasars: absorption lines.
\end{keywords}

%%%%%%%%%%%%%%%%%%%%%%%%%%%%%%%%%%%%%%%%%%%%%%%%%%%%%%%%%%%%%%%%%%%%%	
%%%%%%%%%%%%%%%%%%%%%%%%%% SECTION 1 %%%%%%%%%%%%%%%%%%%%%%%%%%%%%%%%
%%%%%%%%%%%%%%%%%%%%%%%%%%%%%%%%%%%%%%%%%%%%%%%%%%%%%%%%%%%%%%%%%%%%%

\section{Introduction}

High resolution quasar absorption line spectroscopy has enabled the
statistical detection of intergalactic metals at cosmic overdensities
as low as $\Delta=\rho/\langle \rho \rangle=1-10$ at $z\simeq 3$.  The
inferred metallicities are around $10^{-2}-10^{-3}Z_{\odot}$, with
\CIV lines associated with a substantial fraction of \Lya forest lines
with $N_{\rm HI} \ge 10^{14.5}\rm~cm^{-2}$
(\citealt{Cowie95,Ellison00,Simcoe04,Dodorico10}).  Additional metal
lines, such as \SiIV (\citealt{Songaila96,Aguirre04}) and \OVI
(\citealt{Schaye00m,Pieri04,Aguirre08}) are also detected in the low
density intergalactic medium (IGM).  Observations indicate metallicity
increases with density (\citealt{Schaye03,Aracil04}) and the
distribution of metals in the IGM is inhomogeneous
(\citealt{Scannapieco06,Pieri06,Schaye07,FechnerRichter09,Martin10}).

A crucial ingredient in many of these studies are the ionisation
corrections which convert the observed ionic abundances to
metallicities.  These corrections depend on the density and
temperature of the IGM, as well as the intensity and spectral shape of
the metagalactic ultra-violet background (UVB).  The former may be
modelled using cosmological hydrodynamical simulations which follow
the enrichment history of the IGM ({\it e.g.}
\citealt{Theuns02b,Cen05,Oppenheimer06,Tescari10}).  The latter is
usually calculated using detailed models for the UVB intensity and
spectral shape
(\citealt{HaardtMadau96,HaardtMadau01,Fardal98,Faucher09}).
Simulations of intergalactic metal enrichment typically assume the
metagalactic ionising radiation field is spatially uniform.  This is
expected to be a reasonable approximation for hydrogen ionising
photons at $z \simeq 3$, when the mean free path of ionising photons
is much larger than the average separation between ionising sources
(\citealt{MeiksinWhite04,Croft04}).

However, the epoch of \HeII reionisation is thought to complete around
$z\simeq 3$, when quasars are numerous enough to provide the hard
photons ($E>4\rm\,Ryd$) needed to doubly ionise helium
(\citealt{FurlanettoOh08,McQuinn09}).  Indeed, observations of the \HI
and \HeII \Lya forest indicate the UVB spectral shape fluctuates
significantly on scales of $4-10$ Mpc at these redshifts, producing a
wide range of values for the \HeII to \HI column density ratio,
$\eta=N_{\rm HeII}/N_{\rm HI}\sim1-10^{3}$
(\citealt{Zheng04,Fechner06,Shull04,Shull10}).  These
fluctuations are thought to arise from the small number of quasars
lying within the short (\HeII ionising photon) mean free path expected
at the tail-end of \HeII reionisation
(\citealt{Fardal98,Bolton06,Furlanetto09}), although small scale
radiative transfer effects (\citealt{Maselli05,Tittley07}),
collisionally ionised gas (\citealt{Muzahid10}) or a significant
number of thermally broadened lines (\citealt{Fechner07}) may also
produce column density ratios with $\eta \leq 10$.

These spatial inhomgeneities in the UVB spectral shape should also
have an impact on the ionisation balance of the metals in the
IGM.\footnote{There is some evidence for a reduction in the number of
  \CIV absorption systems in close proximity ($\leq 0.3\rm\,Mpc$) to
  quasars at $1.6<z<4$ (\citealt{Wild08}, see also
  \citealt{Fox08,Tytler09b}).  This line-of-sight proximity effect
  arises because of the $1/R^{2}$ dependence of the ionisation
  radiation {\it intensity} around the quasar when the IGM is
  optically thin, and it will manifest itself even after \HeII
  reionisation has long completed.  This effect is slightly different
  to the large scale fluctuations in the UVB {\it spectral shape}
  discussed here, which are expected to be important toward the
  tail-end of \HeII reionisation when the \HeII ionising photon mean
  free path is small.}  Intriguingly, independent of the spectral
shape of the (spatially uniform) UVB used to make ionisation
corrections, several studies find evidence for scatter in the IGM
metallicity at fixed density (\citealt{Rauch97b,Schaye03,Simcoe04}).
This scatter may be intrinsic, or it may instead be partially
attributable to spatial fluctuations in the spectral shape of the UVB.
On the other hand, \cite{Aguirre04} find little evidence for
inhomogeneity in the $[\rm Si/C]$ they infer from the observed optical
depth ratio $\tau_{\rm SiIV}/\tau_{\rm CIV}$, implying that spatial
variations in the UVB spectral shape are small.  There is also some
evidence from metal line analyses for a hardening of the UVB spectrum
at $z\simeq 3$
(\citealt{Savaglio97,Songaila98,Songaila05,Vladilo03,Agafonova05,Agafonova07}),
perhaps due to the onset of \HeII reionisation.  However, other
studies are consistent with no evolution at the same redshifts
(\citealt{Kim02,Boksenberg03,Aguirre04}).  Quantifying the effect of a
spatially inhomogenous UVB on intergalactic metal absorption lines is
thus desirable, both as a potential probe of \HeII reionisation and
for examining the systematic bias, if any, spatially uniform
ionisation corrections impart to constraints on the distribution of
metals in the IGM.

Analytical arguments have previously highlighted the importance of
small scale fluctuations in the hydrogen ionising radiation field for
metals associated with high \HI column density systems
(\citealt{Miralda05,Schaye06}).  However, there has been no detailed
examination of the impact of a spatially inhomogeneous UVB at
$E>4\rm\,Ryd$ on metals associated with lower \HI column densities.  A
self-consistent treatment would require multi-frequency cosmological
radiative transfer in a large ($\ga (200\rm\,Mpc)^{3}$ ) volume
coupled to a detailed chemodynamical model.  In contrast, current
state-of-the-art hydrodynamical simulations of IGM metal enrichment
generally assume a spatially uniform UVB ({\it e.g.}
\citealt{Tescari10,Shen10,CenChisari10,Wiersma10}, but see
\citealt{Oppenheimer09}) while large cosmological radiative transfer
simulations of \HeII reionisation are still restricted to a treatment
of hydrogen and helium only (\citealt{Paschos07,McQuinn09}).

In this paper we instead use a comparatively simple model which
includes most of the relevant aspects of the spatially inhomogeneous
UVB expected towards the tail-end of \HeII reionisation ({\it e.g.}
\citealt{Fardal98,Bolton06,Furlanetto09}).  We combine this model with
a large hydrodynamical simulation of the IGM to explore the impact of
spatial fluctuations in the UVB spectral shape on intergalactic carbon
and silicon at $z=3$.  In particular, we focus on the ratio of \SiIV
to \CIV absorption, which is sensitive to the shape of the UVB either
side of the \HeII ionisation edge
(\citealt{Songaila95,Savaglio97,GirouxShull97}).  We describe our
hydrodynamical simulation and model for spatial fluctuations in the
UVB spectral shape in section 2.  Section 3 describes the four toy UVB
models we use in this work.  In section 4 we proceed to examine the
impact of spatial fluctuations in the UVB spectral shape on synthetic
absorption spectra constructed from our simulations.  We compare our
results to observational data in section 5, and conclude in section 6.
All atomic data is taken from \cite{Morton03}, solar abundances are
from \cite{Asplund09} and all distances are given in comoving units.

%%%%%%%%%%%%%%%%%%%%%%%%%%%%%%%%%%%%%%%%%%%%%%%%%%%%%%%%%%%%%%%%%%%%%	
%%%%%%%%%%%%%%%%%%%%%%%%%% SECTION 2 %%%%%%%%%%%%%%%%%%%%%%%%%%%%%%%%
%%%%%%%%%%%%%%%%%%%%%%%%%%%%%%%%%%%%%%%%%%%%%%%%%%%%%%%%%%%%%%%%%%%%%

\section{Numerical model} \label{sec:numerical}
\subsection{Hydrodynamical simulation of the IGM}

We use the R4 cosmological hydrodynamical simulation described in
\cite{Becker10} to model the IGM density field.  The simulation was
performed in a $40h^{-1}$ Mpc box with $2 \times 512^{3}$ particles
using the parallel Tree-SPH code {\small GADGET-3}
(\citealt{Springel05}).  The cosmological parameters are $\Omega_{\rm
  m}=0.26$, $\Omega_{\Lambda}=0.74$, $\Omega_{\rm b}h^{2}=0.023$,
$h=0.72$, $\sigma_{8}=0.80$, $n_{\rm s}=0.96$ ({\it e.g.}
\citealt{Komatsu09,Reichardt09}).  We use a snapshot drawn from the
simulation at $z=2.976$.

\subsection{A simple model for spatial fluctuations in the UV background spectral shape}

Our model for spatial fluctuations in the UVB spectral shape is
similar to the approach described by \cite{Bolton06} (see also
\citealt{Fardal98,Furlanetto09}).  The model applies to fluctuations
in the UVB spectrum at the tail-end of \HeII reionisation only,
following the overlap of \HeIII regions when the \HeII ionising photon
mean free path is set by the abundance of Lyman limit systems
(\citealt{MiraldaEscude00}).  There is good evidence to suggest that
\HeII reioniation is indeed completing by $z\simeq 3$
(\citealt{Shull10,Becker10}), and this assumption should therefore be
reasonable.  Note, however, that the model will not apply to the
patchy ionisation state of the pre-overlap IGM during the early stages
of \HeII reionisation at $z>3$, and does not include the effect of
detailed, small scale radiative transfer effects on the IGM
(\citealt{Maselli05,Tittley07}).  Further modelling will thus be
required to address the impact of spatial variations in the UVB
spectral shape on intergalactic metals during the heart of \HeII
reionisation.

Firstly, to create a volume large enough to contain several \HeII
ionising photon mean free paths, we stack the hydrodynamical
simulation volume around the original $40h^{-1}$ Mpc box to create a
cube $600h^{-1}\rm\,Mpc$ on each side.  We identify haloes in the
volume using a friends-of-friends halo finder.  Quasar luminosities
are assigned to the haloes by Monte Carlo sampling the
\cite{Hopkins07} B-band quasar luminosity function at $z=3$.  The
luminosities are uniquely mapped to the halo masses by rank ordering
both quantities and assigning the brightest quasars to the most
massive haloes.  This approach enables us to distribute sources on
large scales while following the ionisation state of gas in detail in
the central $40h^{-1}$ Mpc box.  However, it will only approximately
model the clustering properties of the sources.  Fortunately, the
rarity of bright quasars means that Poisson fluctuations rather than
clustering is expected to dominate the resulting UVB fluctuations
(\citealt{Furlanetto09}), so this limitation is unlikely to
significantly affect our results.

We require the number of quasars in a volume, V, to satisfy
 
\begin{equation} N(L_{\rm B}>L_{\rm min}) = V \int_{L_{\rm min}}^{\infty} \phi(L_{\rm B})dL_{\rm B}, \end{equation}

\noindent
where $\phi(L_{\rm B})$ is the \cite{Hopkins07} quasar luminosity
function.  The lower limit of the integral is $L_{\rm
  min}=10^{43.5}\rm \,erg\,s^{-1}$, corresponding to $M_{\rm B}=-20$.
The B-band luminosity of each quasar is converted into a luminosity
below the \HI Lyman limit using a broken power law spectrum ({\it
  e.g.} \citealt{Madau99})

\begin{equation}
L_{\nu}\propto \cases{\nu^{-0.3} &($2500<\lambda<4600\,$\AA),\cr
  \noalign{\vskip3pt}\nu^{-0.8} &($1050<\lambda<2500\,$\AA),\cr
  \noalign{\vskip3pt}\nu^{-\alpha_{\rm s}} &($\lambda<1050\,$\AA).\cr}
\end{equation}

\noindent
We consider two cases for the extreme UV (EUV) spectral index,
$\alpha_{\rm s}$.  In the first case we assume a constant value of
$\alpha_{\rm s}=1.5$, consistent with the mean obtained by
\cite{Telfer02} for radio quiet quasars, $\langle \alpha_{\rm
  s}\rangle=1.57\pm0.17$.  However, a wide range of values are
measured for the EUV spectral index ({\it e.g.}
\citealt{Zheng97,Telfer02,Scott04}). Consequently, we also consider a
variable EUV spectral index. Scatter in $\alpha_{\rm s}$ is included
by Monte-Carlo sampling a Gaussian with mean $1.5$ and standard
deviation $0.5$ over the range $0\leq \alpha_{\rm s} \leq 3$, similar
to the dispersion measured by \cite{Telfer02}.

Once the luminosity and spectral energy distribution for each quasar
is specified, the specific intensity of the UVB, $J({\bf r},\nu)$
$[\rm erg\,s^{-1}\,cm^{-2}\,Hz^{-1}\,sr^{-1}]$, at frequencies between
the \HI and \HeII ionisation edges, $\nu_{\rm HI}< \nu < \nu_{\rm
  HeII}$, is given by

\begin{equation} J({\bf r},\nu) = \frac{1}{4\pi}\sum_{i=1}^{N}\frac{L_{\rm i}({\bf r}_{i},\nu)}{4\pi|{\bf r}_{\rm i}-{\bf r}|^{2}}. \label{eq:JH1} \end{equation} 
 
\noindent 
The sum is made over all quasars in the $600h^{-1}\rm\,Mpc$ volume,
where $|{\bf r}_{\rm i} - {\bf r}|$ is the distance of quasar $i$ from
${\bf r}$, which is always in the central $40h^{-1} \rm Mpc$ box.
Eq.~(\ref{eq:JH1}) assumes that the IGM is optically thin to \HI
ionising photons; this should be a reasonable approximation at $z=3$.

\begin{table}
  \centering
  \caption{The four UVB models used in this work.  From left to right,
    the columns list the model name, the photon mean free path at the
    \HeII ionisation edge, the quasar EUV spectral index and the
    logarithm of the spatially averaged softness parameter, $\log
    \langle S({\bf r}) \rangle = \log[\langle \Gamma_{\rm HI}({\bf
        r})/\Gamma_{\rm HeII}({\bf r}) \rangle]$.}
    \begin{tabular}{c|c|c|c}
      \hline
      Model   & $\lambda_{\rm HeII}$ & $\alpha_{\rm s}$ & log$\langle S({\bf r}) \rangle $ \\
   \hline
    UVB1 &  30 Mpc  & 1.5      & 2.66  \\
    UVB2 &  15 Mpc  & 1.5      & 3.10  \\
    UVB3 &  30 Mpc  & Variable & 2.51  \\
    UVB4 &  30 Mpc  & 1.5      & 2.66  \\
    \hline
\end{tabular}
\label{tab:sims}

\end{table}

\begin{figure*}
\centering
\begin{minipage}{180mm}
\begin{center}
   \psfig{figure=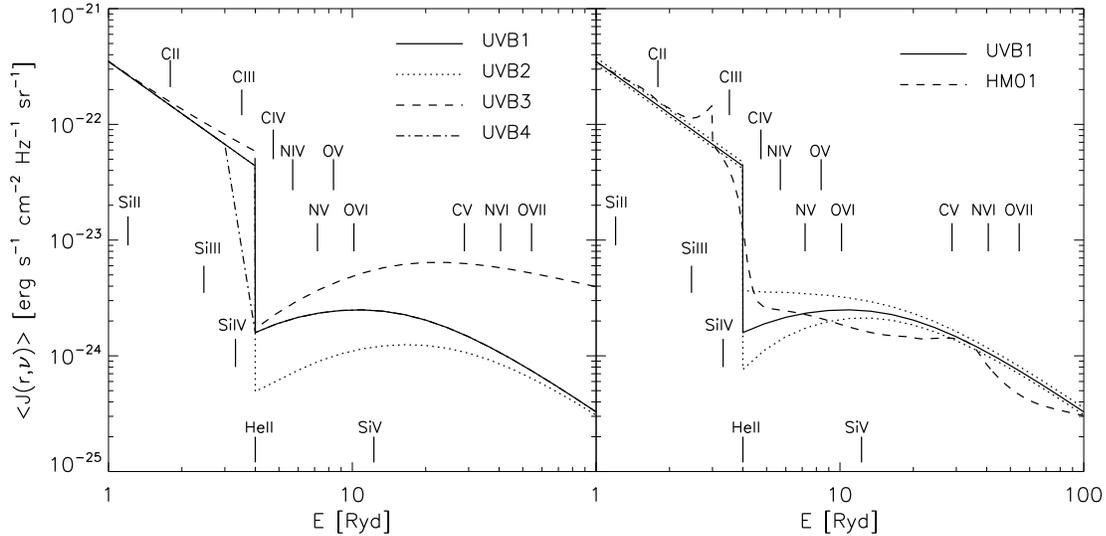,width=0.9\textwidth}
    
\vspace{-0.4cm}
\caption{{\it Left:} The spatially averaged specific intensity as a
  function of photon energy, $E=h\nu$, for the four spatially
  fluctuating UVB models summarised in Table 1.  Models UVB1 (solid
  curve) and UVB4 (dot-dashed curve) are identical at $E>4\rm\,Ryd$.
  Models UVB1 and UVB2 (dotted curve) are the same at $E<4\rm\,Ryd$.
  Vertical tick marks correspond to the ionisation potentials of
  various ions. {\it Right:} The solid curve displays the spatially
  averaged specific intensity for model UVB1.  The range in specific
  intensity, $J({\bf r},\nu)$, encompassing 95 per cent of all
  fluctuations from the median is displayed by the dotted curves.  For
  comparison, the dashed line shows the UVB background model of
  \citet{HaardtMadau01} for emission from galaxies and quasars at
  $z=3$, renormalised to have specific intensity $J(\nu_{\rm
    HI})=3.5\times
  10^{-22}\rm\,erg\,s^{-1}\,cm^{-2}\,Hz^{-1}\,sr^{-1}$.}

\label{fig:UVBspectra}
\end{center}
\end{minipage}
\end{figure*}

At frequencies above the \HeII ionisation edge, $\nu>\nu_{\rm HeII}$,
the specific intensity is instead

\begin{equation} J({\bf r},\nu) = \frac{1}{4\pi}\sum_{i=1}^{N}\frac{L_{i}({\bf r}_{\rm i},\nu)}{4\pi|{\bf r}_{\rm i}-{\bf r}|^{2}}e^{-\frac{ |{\bf r}_{\rm i}-{\bf r}|}{\lambda_{\rm HeII}}\left(\frac{\nu}{\nu_{\rm HeII}}\right)^{-3(\beta-1)}}, \label{eq:JHe2} \end{equation} 

\noindent
where $\lambda_{\rm HeII}$ is the mean free path for photons at the
\HeII ionisation edge and $\beta$ is the power-law index describing
the slope of the \HeII column density distribution, $f(N_{\rm HeII})
\propto N_{\rm HeII}^{-\beta}$.  This expression assumes quasars are
point sources surrounded by discrete, Poisson distributed absorbers
(\citealt{Faucher09}).  We adopt $\beta=1.5$, following the analogous
distribution for \HI absorbers (\citealt{Petitjean93,Kim02}).  Note,
however, the slope of the \HeII column density distribution is not
constrained observationally and may deviate from the \HI distribution
for \HeII absorbers associated with high \HI column densities, $N_{\rm
  HI} \ga 10^{16}\rm\,cm^{-2}$ (\citealt{Fardal98,Faucher09}).
Finally, in this work we model the contribution to the UVB from
quasars only.  The non-thermal emission from quasars is thought to
dominate the UVB at $\nu \ga \nu_{\rm HeII}$, but star forming
galaxies will make an increasingly significant contribution to the UVB
at lower frequencies at $z>3$
(\citealt{Bianchi01,Haehnelt01,Bolton05,Kirkman05,Faucher08b}).
However, the soft spectra of these sources are unlikely to
significantly contribute to spatial fluctuations in the UVB spectral
shape at $E>4\rm\,Ryd$.

%%%%%%%%%%%%%%%%%%%%%%%%%%%%%%%%%%%%%%%%%%%%%%%%%%%%%%%%%%%%%%%%%%%%%	
%%%%%%%%%%%%%%%%%%%%%%%%%% SECTION 3 %%%%%%%%%%%%%%%%%%%%%%%%%%%%%%%%
%%%%%%%%%%%%%%%%%%%%%%%%%%%%%%%%%%%%%%%%%%%%%%%%%%%%%%%%%%%%%%%%%%%%%

\section{Toy UV background spectra}

The four toy UVB models used in this work are summarised in Table 1.
The models are constructed by evaluating Eqs.~(\ref{eq:JH1}) and
(\ref{eq:JHe2}) on a $14^3$ grid within the central $40h^{-1}\rm\,Mpc$
simulation box.  The size of each grid cell, $\sim 4$ Mpc, is chosen
to match the scales of $4-10$ Mpc on which the \HeII to \HI optical
depth ratio is observed to vary at $z=2.4-2.9$ (\citealt{Shull10}).
This choice also represents the best compromise between resolution and
efficiency; the construction of the photo-ionisation balance look-up
tables (see section~\ref{sec:makespec}) becomes very time consuming if
the number of grid cells much larger.  However, note that the rarest,
high amplitude fluctuations in the UVB spectral shape on small scales
will not be fully captured in the simulation.  In each grid cell the
specific intensities are evaluated at $25$ different photon energies
spanning the range $E=1-100\rm\,Ryd$.  Lastly, all four models are
renormalised by a factor of $0.74$ to give identical mean specific
intensities at the \HI ionisation edge $\langle J({\bf r},\nu_{\rm
  HI})\rangle=3.5\times
10^{-22}\rm\,erg\,s^{-1}\,cm^{-2}\,Hz^{-1}\,sr^{-1}$.  This
renormalisation yields \HI photo-ionisation rates of $\Gamma_{\rm HI}
\sim 1.0 \times 10^{-12}\rm\,s^{-1}$ (cf. $\Gamma_{\rm HI} \sim
1.3\times 10^{-12}\rm\,s^{-1}$ prior to the renormalisation),
consistent with observational constraints derived from the \Lya forest
opacity at $z=3$ (\citealt{Bolton05,Faucher08b}).

Models UVB1 and UVB2 are constructed using a single EUV spectra index,
$\alpha_{\rm s}=1.5$.  However, the value of the mean free path at the
\HeII ionisation edge, $\lambda_{\rm HeII}$, is rather uncertain.  In
this work we adopt a fiducial value of $\lambda_{\rm HeII}=30\rm\,Mpc$
at $z=3$ ({\it e.g.}  \citealt{Fardal98,MiraldaEscude00,Bolton06}),
but the exact value will depend on the \HeII to \HI column density
ratio, $\eta=N_{\rm HeII}/N_{\rm HI}$, which exhibits significant
fluctuations approaching $z=3$ (\citealt{Zheng04,Shull04,Shull10}).
We thus also consider a smaller mean free path for model UVB2,
$\lambda_{\rm HeII}=15\rm\,Mpc$, consistent with the constraints
recently presented by \cite{DixonFurlanetto09} at $z\simeq 3$.  Model
UVB3 instead uses a variable EUV spectral index and $\lambda_{\rm
  HeII}=30$ Mpc.  The final model, UVB4, is identical to model UVB1,
aside from the spectral shape between $3$ and $4$ Ryd.  Resonant
absorption due to \HeII Lyman series lines will also produce
significant spatial fluctuations in the UVB between \HeII \Lya at
$3\rm\,Ryd$ and the \HeII ionisation edge at $4\rm\,Ryd$
(\citealt{MadauHaardt09}).  Model UVB4 mimics these fluctuations by
assuming the UVB spectrum computed with Eqs.~(\ref{eq:JH1}) and
(\ref{eq:JHe2}) follows a power law $J({\bf r},\nu)=J({\bf r},\nu_{\rm
  Ly\alpha})(\nu/\nu_{\rm Ly\alpha})^{\xi}$ for $\nu_{\rm Ly\alpha}<
\nu < \nu_{\rm HeII}$, where $\nu_{\rm Ly\alpha}$ is the frequency of
\HeII \Lya at $3\rm\,Ryd$ and $\xi \simeq 8\log[J({\bf r},\nu_{\rm
    HeII})/J({\bf r},\nu_{\rm Ly\alpha})]$.

\begin{figure*}
\centering
\begin{minipage}{180mm}
\begin{center}
   \psfig{figure=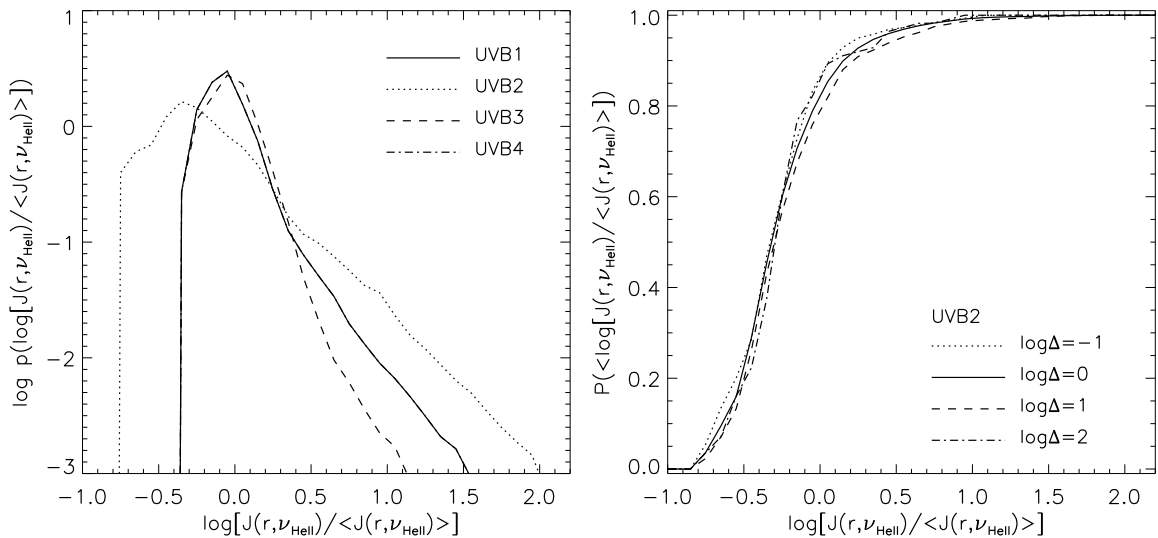,width=0.9\textwidth}

\vspace{-0.4cm}

\caption{{\it Left:} The probability distribution for fluctuations in
  the specific intensity, $J({\bf r},\nu_{\rm HeII})$, relative to the
  spatially averaged value in the simulation volume, $\langle J({\bf
    r},\nu_{\rm HeII}) \rangle$, for the four UVB models summarised in
  Table 1.  The distributions are calculated using the specific
  intensity at the \HeII ionisation edge, where the fluctuations are
  largest.  Note that the distributions for models UVB1 and UVB4 are
  identical at this frequency. {\it Right:} Cumulative probability
  distributions for fluctuations in the specific intensity of UVB2 at
  $4\rm\,Ryd$.  The four curves show the cumulative distributions
  obtained in pixel subsets corresponding to overdensities at $\log
  \Delta=-1,\,0,\,1$ and $2$.}
\label{fig:flucPDF}
\end{center}
\end{minipage}
\end{figure*}

The spatially averaged spectra, $\langle J({\bf r},\nu) \rangle$, for
all four UVB models are displayed in the left hand panel of
Fig.~\ref{fig:UVBspectra}. The step at $4\rm\,Ryd$ in models UVB1,
UVB2 and UVB3, and between $3-4\rm\,Ryd$ for model UVB4, is due to the
attenuation of the spectrum by the \HeII opacity.  The spectra recover
their intrinsic shape at high energies, where the photon mean free
path is long and the attenuation of the radiation by the intervening
\HeII absorbers is negligible.  Model UVB3, which assumes a variable
EUV spectral index, is much harder than the other models at high
energies.  This model is dominated by the quasars with the hardest
spectra, $\alpha_{\rm s} \sim 0$, in the simulation volume, even if
these objects are not the brightest quasars at lower frequencies.  The
logarithm of the spatially averaged softness parameter for each model,
defined as the ratio of the \HI to \HeII photo-ionisation rates,
$S=\Gamma_{\rm HI}/\Gamma_{\rm HeII}$, are listed in Table 1.
Observational constraints on the softness parameter at $z=3$ are
dominated by the large uncertainty in the \HeII effective optical
depth.  \cite{Bolton06} find $2.3 \leq \log S \leq \ 3.2$ at $z=3$ by
comparing hydrodynamical simulations of the \HI and \HeII \Lya forest
opacity to observational data.  All four models are designed to be
consistent with these constraints.

The right hand panel of Fig.~\ref{fig:UVBspectra} again displays model
UVB1, but the dotted curves now show the range encompassing 95 per
cent of all fluctuations around the median specific intensity.
Fluctuations are largest close to the \HeII ionisation edge where the
attenuation of the radiation is strongest, while towards higher
frequencies the increasing mean free path produces much smaller
departures from the median.  For comparison, the dashed line shows the
widely used UVB model of \citet{HaardtMadau01} (HM01) for emission
from galaxies and quasars at $z=3$.  The spectrum has been
renormalised to match the specific intensity of UVB1 at $1\rm\,Ryd$.
The softness parameter of the HM01 spectrum is $\log S = 2.42$, around
$0.2\rm\, dex$ lower than UVB1.  Although the features of the toy UVB
spectra presented here are broadly similar to the more detailed model
of HM01, there are some important differences.  In particular, our toy
models do not include more complex effects such as recombination
emission (\citealt{HaardtMadau96,Fardal98,Faucher09}) or sawtooth
absorption by the \HeII Lyman series (\citealt{MadauHaardt09}) which
are important to consider when performing detailed modelling of
observational data.

The fluctuations in the specific intensity are quantified in more
detail in Fig.~\ref{fig:flucPDF} for all four UVB models.  The left
panel displays the probability distribution for the specific intensity
relative to its spatially averaged value, $\log[J({\bf r},\nu_{\rm
  HeII})/\langle J({\bf r},\nu_{\rm HeII}) \rangle]$, at the \HeII
ionisation edge.  The distribution becomes broader for a smaller
$\lambda_{\rm HeII}$ (UVB2) as the number of sources within one mean
free path is reduced.  Somewhat counter-intuitively, the variable EUV
spectral index model (UVB3) produces fewer of the large, rare
fluctuations from the mean than UVB1, which assumes a constant
$\alpha_{\rm s}$ but is otherwise identical.  This is because the
three quasars which lie closest to the centre of our simulation volume
(and hence dominate the rare, high amplitude fluctuations) have
randomly assigned EUV spectral indices which conspire to reduce
fluctuations in the specific intensity at $4\rm\,Ryd$.  Recall the
quasar luminosities are normalised in the B-band, and in this instance
the EUV spectra for the three closest quasars all have $\alpha_{\rm
  s}>1.5$.  This produces lower specific intensities at the \HeII
ionisation edge compared to model UVB1 with $\alpha_{\rm s}=1.5$, and
hence fewer large fluctuations from the mean.

Lastly, the right panel of Fig.~\ref{fig:flucPDF} displays cumulative
distributions for $\log[J({\bf r},\nu_{\rm HeII})/\langle J({\bf
    r},\nu_{\rm HeII}) \rangle]$ obtained from model UVB2.  The four
separate curves show the cumulative distributions for subsets of
pixels corresponding to regions in the IGM density field with
overdensities at $\log \Delta=-1,\,0,\,1$ and $2$, respectively.  The
lack of a clear correlation with density
\footnote{\cite{Bolton06} found that the \HeII to \HI column density
  ratio, $\eta=N_{\rm HeII}/N_{\rm HI}$, obtained from a Voigt profile
  analysis of synthetic absorption spectra was systematically higher
  in regions where $\tau_{\rm HI}>0.05$ compared to regions with lower
  \HI optical depths (and hence densities).  They suggested this
  implied the fluctuating UVB model they used was harder in higher
  density regions.  However, this interpretation is in disagreement
  with our findings here.  This earlier result is likely spurious, and
  probably results from the assumptions of pure turbulent broadening
  in the Voigt profile analysis (\citealt{Fechner07}) and a spatially
  uniform \HI photo-ionisation rate.} is mainly due the low number
density of bright quasars, resulting in a radiation field which varies
on scales much larger than the Jeans scale (see the discussions in
{\it e.g.}  \citealt{McQuinn09,Furlanetto09b}).  Note, however, that
fluctuations on scales $<4\rm\,Mpc$ (typically those associated with
the radiation field in close proximity to the quasars) will be
smoothed over, but these fluctuations are also rather rare. On the
other hand, if additional physics such as small scale radiative
transfer effects are important, or if \HeII ionising sources are far
more numerous ({\it e.g.}  galaxies), the ionising radiation field may
be more closely correlated with the IGM density on small scales.

%%%%%%%%%%%%%%%%%%%%%%%%%%%%%%%%%%%%%%%%%%%%%%%%%%%%%%%%%%%%%%%%%%%%%	
%%%%%%%%%%%%%%%%%%%%%%%%%% SECTION 4 %%%%%%%%%%%%%%%%%%%%%%%%%%%%%%%%
%%%%%%%%%%%%%%%%%%%%%%%%%%%%%%%%%%%%%%%%%%%%%%%%%%%%%%%%%%%%%%%%%%%%%

\section{The impact of UVB fluctuations on simulated absorption spectra}
\subsection{Construction of synthetic absorption spectra} \label{sec:makespec}

\begin{figure*}
\centering
\begin{minipage}{180mm}
\begin{center}
   \psfig{figure=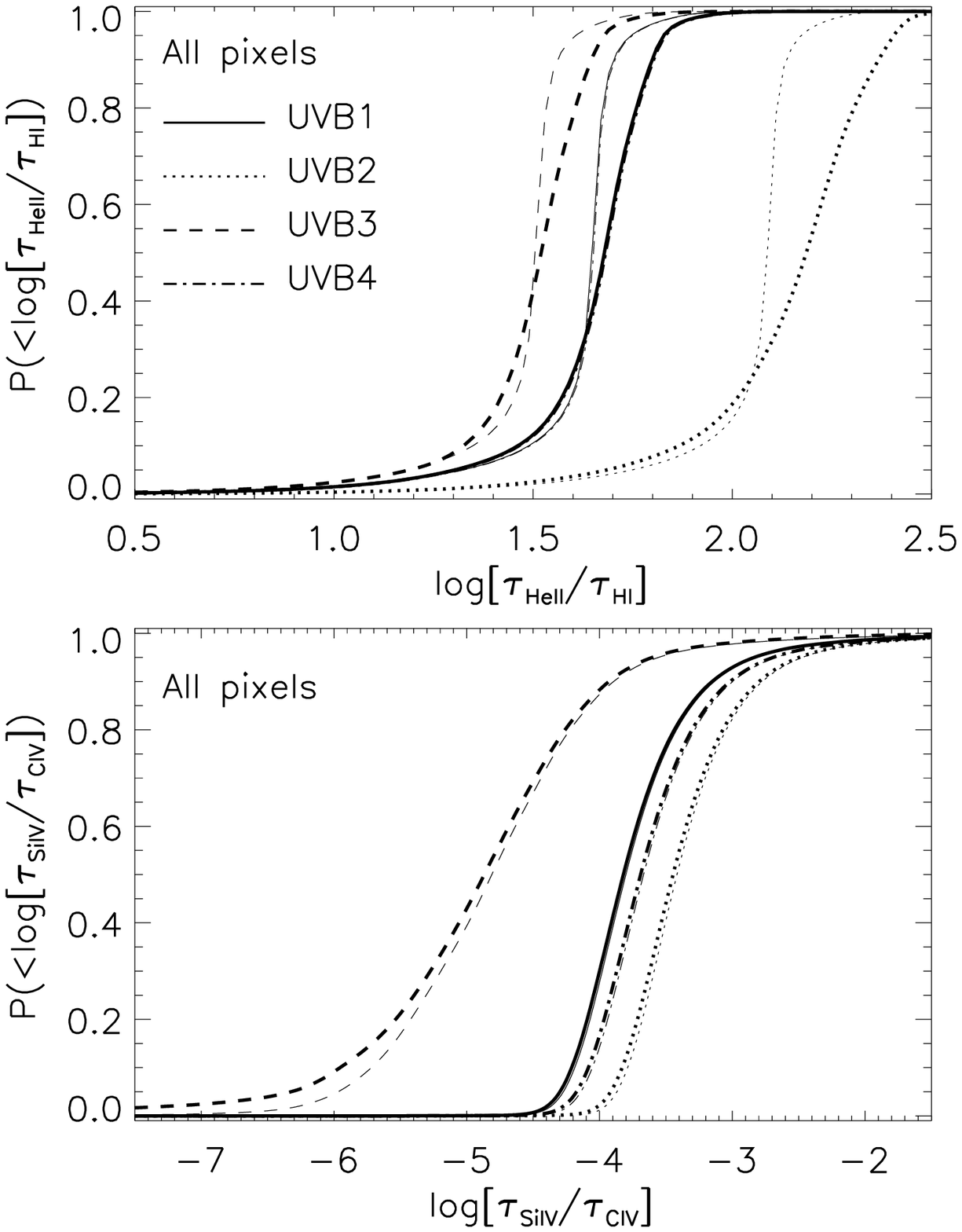,width=0.45\textwidth}
   \psfig{figure=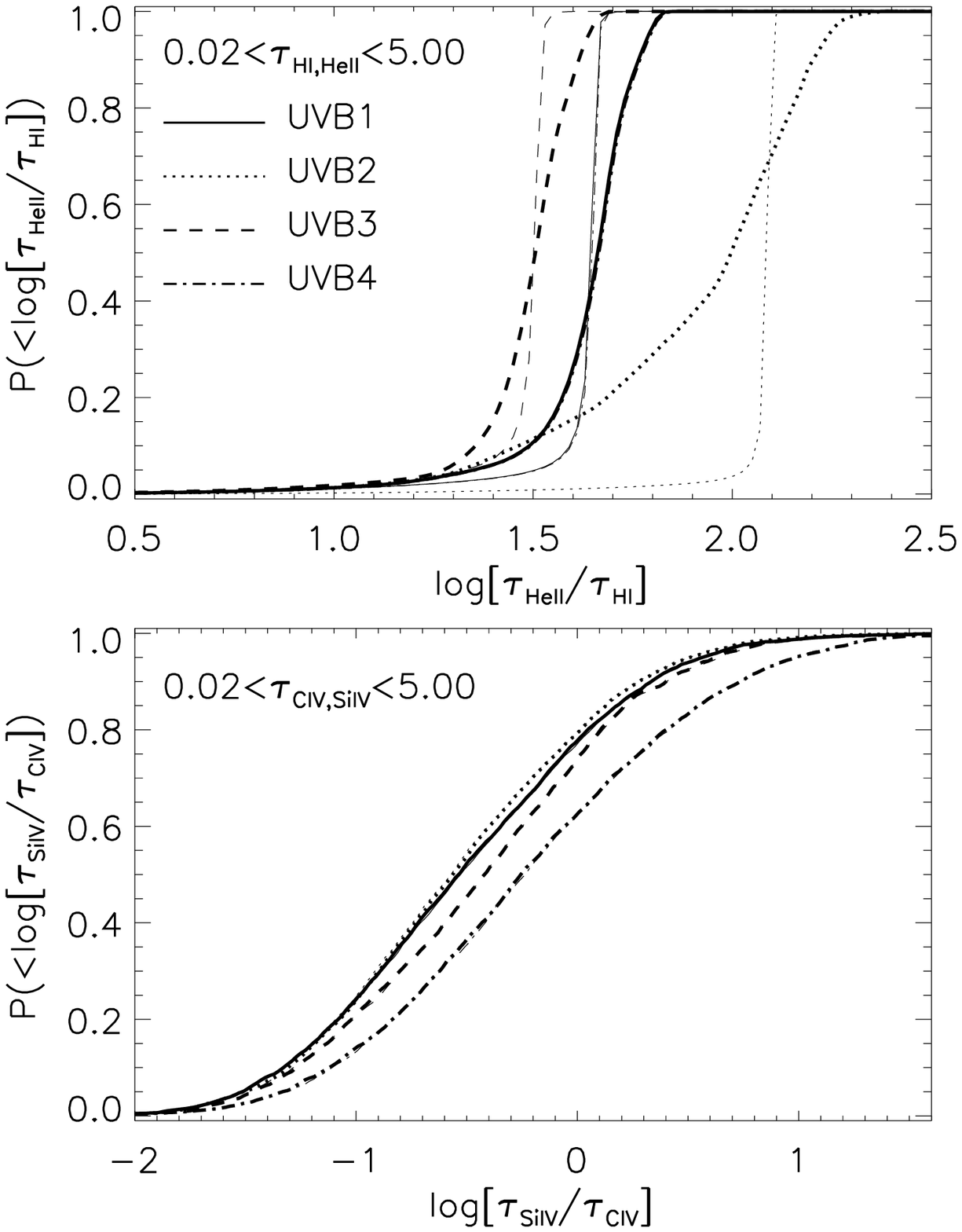,width=0.45\textwidth}
   \vspace{-0.3cm}
  \caption{{\it Upper left:} The cumulative probability distribution
    for $\tau_{\rm HeII}/\tau_{\rm HI}$ for {\it all pixels} in the
    sight-lines drawn from our hydrodynamical simulation.  The curves
    correspond to models UVB1 (solid), UVB2 (dotted), UVB3 (dashed)
    and UVB4 (dot-dashed).  Thick curves are obtained from spectra
    constructed using the spatially fluctuating UVB.  Thin curves
    correspond to the optical depth ratios obtained using the
    spatially averaged UVB spectrum. {\it Upper right:} As for the
    upper left panel, but now for pixels with $0.02<\tau_{\rm
      HI,HeII}<5$ only.  Models UVB1 and UVB4 are indistinguishable in
    both of these panels.  {\it Lower left:} The cumulative
    probability distribution for $\tau_{\rm SiIV}/\tau_{\rm CIV}$ for
    all pixels.  {\it Lower right:} As in the lower left panel, but
    now for pixels with $0.02<\tau_{\rm CIV,SiIV}<5$ only.  The thin
    and thick curves are almost indistinguishable in the lower panels.
    Note also the scales on the horizontal axes of the lower left and
    right hand panels are significantly different.}
\label{fig:ratios}

\end{center}
\end{minipage}
\end{figure*}

We now turn to examine the impact of our spatially inhomogeneous UVB
models on synthetic quasar absorption spectra.  We use a simple {\it a
  posteriori} approach to including metals within our hydrodynamical
simulation.  We do not model the production and dispersal of metals in
detail (see {\it e.g.}
\citealt{Theuns02b,Cen05,Oppenheimer06,Shen10,Wiersma10,Tescari10}).
We instead follow carbon and silicon abundances only, and assume the
metallicity of the IGM traces the gas density
distribution. \cite{Schaye03} measure a best fit median metallicity of
$\rm [C/H]=-3.47^{+0.07}_{-0.06} + 0.65^{+0.10}_{-0.14}(\log \Delta -
0.5)$, with a lognormal scatter of
$\sigma({\rm [C/H]})=0.76^{+0.05}_{-0.08}-0.23^{+0.09}_{-0.07}(\log\Delta -
0.5)$ at $z=3$.  Following \cite{Schaye03}, we introduce the lognormal
scatter to the median as ${\rm [C/H]}=-3.47 + 0.65(\log \Delta - 0.5)
+s$ by dividing the $40h^{-1}$ Mpc simulation volume into $32^{3}$
cubes and adding a different scatter, $s$, in each sub-volume.  The
scatter, $s$, is obtained by randomly sampling a normal distribution
in $[\rm C/H]$ with zero mean and standard deviation $\sigma=0.76$.
We further assume the silicon abundance relative to carbon is $\rm
[Si/C]=0.77$ (\citealt{Aguirre04}).  Note, however, these metallicity
constraints are derived from a statistical analysis of \CIV and \SiIV
absorption lines which use an ionisation correction based on the HM01
UVB model.  Different ionisation corrections obtained using
significantly softer or harder UVB spectra will alter these
constraints considerably.

The equilibrium ionisation balance of the metals in our hydrodynamical
simulation is calculated using the photo-ionisation code Cloudy
(version 08.00), last described by \cite{Ferland98}.  For each of our
four spatially inhomogeneous UVB models, we create a five dimensional
look-up table listing the ionisation fractions of hydrogen, helium,
carbon and silicon as a function of gas temperature, $T$, hydrogen
number density, $n_{\rm H}$, and position on the $14^{3}$ UVB grid
within our $40h^{-1}\rm\,Mpc$ simulation box.  Synthetic spectra are
obtained from the hydrodynamical simulations by randomly selecting
1000 sight-lines from the simulation and performing an interpolation
on the particle data weighted by the smoothing kernel ({\it e.g.}
\citealt{Theuns98}).  Each line of sight is drawn parallel to the box
boundaries and has 2048 pixels, each of which has an associated gas
density, temperature, peculiar velocity and ionisation fraction.  The
ionisation fractions for H$\,\rm \scriptstyle I$, He$\,\rm
\scriptstyle II$, C$\,\rm \scriptstyle IV$ and Si$\,\rm \scriptstyle
IV$ are obtained by linearly interpolating the five dimensional
look-up table.  Absorption spectra are constructed for \HI \Lya
($\lambda$1216), \Lyb ($\lambda$1026), \Lyg ($\lambda$973), \Lyd
($\lambda$950), \HeII \Lya ($\lambda$304), \CIV ($\lambda\lambda$1548,
1551) and \SiIV ($\lambda\lambda$1394, 1403).

\subsection{Optical depth ratios}

The ratio of the optical depths for two different species is sensitive
to fluctuations in the spectral shape of the ionising background
between their respective ionisation potentials.  A ratio also has the
advantage of being less susceptible to fluctuations in the IGM density
field (\citealt{Worseck07,FurlanettoLidz10}).  We examine $\tau_{\rm
  HeII}/\tau_{\rm HI}$ for the \Lya transitions and $\tau_{\rm
  SiIV}/\tau_{\rm CIV}$ for $\lambda$1548 and $\lambda$1394 in our
synthetic spectra.  Both of these ratios have been widely used as
probes of \HeII reionisation and the UVB spectral shape at $z\simeq 3$
({\it e.g.}  \citealt{Songaila98,Schaye03,Worseck07,Shull10}).  Note
the latter will also be sensitive to fluctuations in the relative
abundance of carbon and silicon.

\subsubsection{The $\tau_{\rm HeII}/\tau_{\rm HI}$ ratio}

It is instructive to first revisit the effect of UVB fluctuations on
the \HeII to \HI optical depth ratio (but see also
\citealt{Fardal98,Maselli05,Bolton06,Tittley07,McQuinn09,Furlanetto09}).
The upper left panel of Fig.~\ref{fig:ratios} shows the cumulative
distribution for $\tau_{\rm HeII}/\tau_{\rm HI}$ for {\it all pixels}
in our synthetic spectra.  This includes pixel optical depths that are
in practice unobservable, either because of saturated absorption or
noise.  The thick curves correspond to the optical depth ratios for
the four spatially fluctuating UVB models listed in
Table~\ref{tab:sims}.  Models UVB1 and UVB4 are almost identical;
neither \HI or \HeII are significantly ionised by photons with
$E=3-4\rm\,Ryd$.  Models with progressively harder spectra produce
lower median values for the \HeII to \HI ratio.  The \HeII fraction
decreases as the specific intensity at $E> 4\rm\,Ryd$ increases; since
the UVB is almost uniform at $1\rm\,Ryd$ in all our models (see the
right panel in Fig.~\ref{fig:UVBspectra}), fluctuations in the
specific intensity at the \HI ionisation edge will have little impact
on $\tau_{\rm HeII}/\tau_{\rm HI}$.  Note also that the smaller mean
free path used for UVB2 produces a much wider range of $\tau_{\rm
  HeII}/\tau_{\rm HI}$ values due to the wider range of fluctuations
produced at the \HeII ionisation edge.

We may assess the extent to which $\tau_{\rm HeII}/\tau_{\rm HI}$ is
influenced by fluctuations in the UVB spectral shape by comparison to
the ratio obtained for a spatially uniform UVB. The thin curves
display the optical depth ratios obtained using the spatially averaged
({\it i.e.} spatially uniform) spectrum for each UVB model.  The
distributions for models including UVB fluctuations are indeed
broader, especially toward higher values of $\tau_{\rm HeII}/\tau_{\rm
  HI}$.  However, all four distributions still extend to low
$\tau_{\rm HeII}/\tau_{\rm HI}$ in the spatially uniform case.  This
behaviour is almost entirely due to variations in $\tau_{\rm
  HeII}/\tau_{\rm HI}$ introduced by thermal broadening; the thermal
width of an absorption line for species $i$ is $b_{\rm i}\propto
m_{\rm i}^{-1/2}$, where $m_{\rm i}$ is the mass of the particle.
Fluctuations in the optical depth ratio therefore also originate from
the wings of absorption lines which are predominantly thermally rather
than turbulently broadened (\citealt{Fechner07}).

The upper right panel of Fig.~\ref{fig:ratios} instead shows only the
subset of pixels which have $0.02<\tau_{\rm HI,HeII}<5$, roughly
corresponding to range over which the optical depths may be reliably
measured in observational data ({\it e.g.} \citealt{Shull10}).  These
pixels correspond to $27.4$, $4.1$, $39.1$ and $27.1$ per cent of the
total for models UVB1, UVB2, UVB3 and UVB4, respectively.  The median
$\tau_{\rm HeII}/\tau_{\rm HI}$ for UVB2 is now significantly smaller.
This is because the small number of optical depths with
$0.02<\tau_{\rm HeII}<5$ selected for UVB2 tend to probe regions where
the \HeII fraction is lowest.

Recent observations of the \HeII \Lya opacity in the spectrum of the
$z\simeq 2.9$ quasar HE2345$-$4342, made using the Cosmic Origins
Spectrograph on the {\it Hubble Space Telescope}, reliably measure
fluctuations in the optical depth ratio spanning the range $\tau_{\rm
  HeII}/\tau_{\rm HI}\simeq 1.25-125$ at $2.75<z<2.89$
(\citealt{Shull10}).  In comparison, model UVB2 exhibits $\tau_{\rm
  HeII}/\tau_{\rm HI}\simeq 0.8-248$ ($15-188$ for 95 per cent around
the median), while UVB1 and UVB4 have $\tau_{\rm HeII}/\tau_{\rm
  HI}\simeq 0.4-77$ ($16-65$) and UVB3 has $\tau_{\rm HeII}/\tau_{\rm
  HI}\simeq 0.3-58$ ($13-46$).  We do not perform a more detailed
comparison here, but as previously noted these results strongly
suggest the observed $\tau_{\rm HeII}/\tau_{\rm HI}$ fluctuations are
largely attributable spatial variations in the \HeII ionising
background.  These fluctuations arise primarily from the small number
of quasars within the short \HeII ionising photon mean free path,
$\lambda_{\rm HeII} \sim 15-30$ Mpc, expected towards the tail-end of
\HeII reionisation.

\subsubsection{The $\tau_{\rm SiIV}/\tau_{\rm CIV}$ ratio}

The results for the $\tau_{\rm SiIV}/\tau_{\rm CIV}$ ratio are shown
in the lower panels of Fig.~\ref{fig:ratios}.  The ionisation edges
for \SiIV and \CIV are at $3.32$ and $4.74$ Ryd respectively; the
ratio of these optical depths is thus sensitive to the shape of the
UVB either side of the \HeII ionisation edge at $4$ Ryd
(\citealt{Songaila95,GirouxShull97}).  The left hand panel displays
the distributions for all pixels in the synthetic spectra.  Models
with progressively harder spectra produce smaller median values for
the ratio -- the distribution for UVB3 in particular extends to
significantly lower $\tau_{\rm SiIV}/\tau_{\rm CIV}$ values.  A harder
UVB reduces both the \SiIV and \CIV ionisation fractions, but the
\SiIV fraction is lowered more rapidly as the break at $4\rm\,Ryd$ is
reduced, leading to the smaller values for the $\tau_{\rm
  SiIV}/\tau_{\rm CIV}$ ratio in UVB3.  However, in contrast to the
case for $\tau_{\rm HeII}/\tau_{\rm HI}$, the distributions for the
fluctuating and uniform UVB models are almost identical.  Only the
UVB3 distribution displays a small difference for the lowest values of
$\tau_{\rm SiIV}/\tau_{\rm CIV}$.  This scatter in the $\tau_{\rm
  SiIV}/\tau_{\rm CIV}$ ratio is instead dominated by the different
dependences of the \CIV and \SiIV fractions on gas density, and to a
much lesser extent variations in the absorption profiles in redshift
space.

The lower right hand panel in Fig.~\ref{fig:ratios} displays the
subset of pixels with $0.02<\tau_{\rm SiIV,CIV}<5$.  These pixels
represent a very small fraction of the total: $0.2$, $0.3$, $0.1$ and
$0.2$ per cent for models UVB1, UVB2, UVB3 and UVB4, respectively,
reflecting the much smaller volume filling factor of detectable \CIV
and \SiIV absorption in the simulations.  The medians of all four
distributions are now significantly larger; most of the discarded
pixels have \CIV and \SiIV optical depths which are too low to be
observed.  The distributions for models UVB1, UVB2 and UVB3 are now
similar, suggesting that the $\tau_{\rm SiIV}/\tau_{\rm CIV}$ ratio is
a less sensitive probe of the UVB spectral shape for these pixels.
Finally, all four distributions are again virtually indistinguishable
from the distributions obtained with a spatially uniform UVB.

\subsection{Gas temperature and density}

\begin{figure}
\begin{center}
  \includegraphics[width=0.48\textwidth]{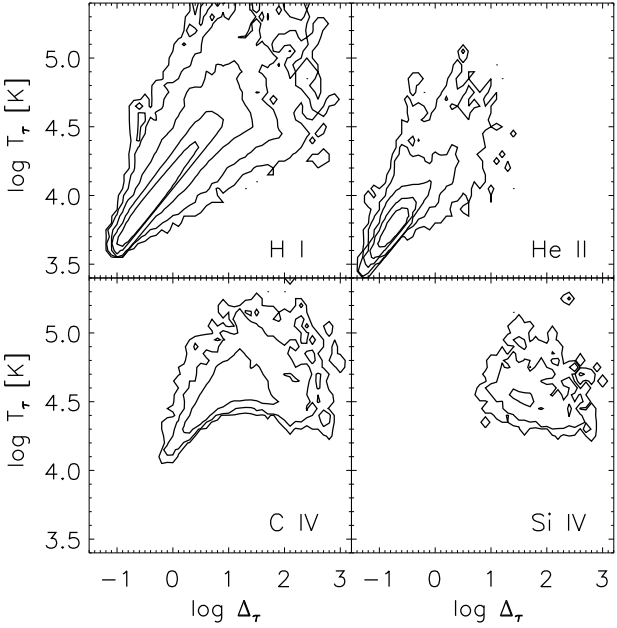}
  \vspace{-0.4cm}
  \caption{Contour plots of the optical depth weighted temperature,
    $T_{\tau}$, against the optical depth weighted overdensity,
    $\Delta_{\tau}$, for pixels with $0.02<\tau <5$ in the synthetic
    absorption spectra constructed using model UVB1.  {\it Top left:}
    The temperature density plane for \HI absorption. {\it Top right:}
    \HeII absorption. {\it Bottom left:} \CIV absorption. {\it Bottom
      right:} \SiIV absorption. The number density of pixels increases
    by 1.0 dex within each contour level.}
\label{fig:Trho}
\end{center}
\end{figure}

We may gain insight into this behaviour by first examining the typical
gas densities and temperatures which produce the absorption in our
synthetic spectra.  Contour plots of the optical depth weighted
temperature against optical depth weighted overdensity are shown in
Fig.~\ref{fig:Trho} for spectra constructed using model UVB1.
Clockwise from the upper left, each panel shows the
temperature-density plane for H$\,\rm \scriptstyle I$, He$\,\rm
\scriptstyle II$, Si$\,\rm \scriptstyle IV$ and C$\,\rm \scriptstyle
IV$ optical depths with $0.02<\tau<5$.  The number of pixels increases
by an order of magnitude within successive contours.  It is evident
that \HI and \HeII absorption primarily probes photo-ionised gas
around mean density and below, with the \HeII absorption slightly more
sensitive to gas in voids (\citealt{Croft97,McQuinn09b}).  In
contrast, \CIV and especially \SiIV tend to probe {\it overdense}
regions in the IGM ({\it e.g.}  \citealt{Rauch97b}).  Nearly all the
carbon and silicon absorption is furthermore associated with gas at
$T<10^{5}\rm\,K$.

The exact temperature and density of regions with measurable optical
depths depends on the UVB model used.  For example, for UVB3 (the
hardest spectrum we consider) the median $\tau_{\rm HI}$ weighted
temperature and overdensity for pixels with $0.02<\tau_{\rm HI, HeII}
<5$ are $T_{\tau}=7\,100\rm\,K$ and $\log \Delta_{\tau}=-0.6$. For
$0.02<\tau_{\rm CIV, SiIV} <5$ the median $\tau_{\rm CIV}$ weighted
values are instead $T_{\tau}=34\,500\rm\,K$ and $\log
\Delta_{\tau}=2.0$. For UVB2 (the softest spectrum) the corresponding
values are $T_{\tau}=5\,600\rm\,K$, $\log \Delta_{\tau}=-0.8$ for
$0.02<\tau_{\rm HI, HeII} <5$ and $T_{\tau}=43\,400\rm\,K$, $\log
\Delta_{\tau}=1.6$ for $0.02<\tau_{\rm CIV, SiIV} <5$.  Nevertheless,
it is clear the $\tau_{\rm HI}/\tau_{\rm HeII}$ and $\tau_{\rm
  SiIV}/\tau_{\rm CIV}$ ratios probe rather different temperature and
density regimes in the IGM.  Furthermore, since the majority of the
\CIV and \SiIV absorption in our models corresponds to gas with
$T<10^{5}\rm\,K$, collisional ionisation is unlikely to explain the
similarity between the $\tau_{\rm SiIV}/\tau_{\rm CIV}$ distribution
for the spatially uniform and fluctuating cases shown in
Fig.~\ref{fig:ratios}.  Note, however, additional heating from
feedback is not included in our hydrodynamical simulation, and we thus
likely underpredict the amount of \CIV and \SiIV absorption due to
collisionally ionised gas.  However, because fluctuations in the
photo-ionising background will be less important for the $\tau_{\rm
  SiIV}/\tau_{\rm CIV}$ ratio if a larger fraction of these elements
are in a hot, predominantly collisionally ionised phase with $\log
\Delta \ga 2$, this is unlikely to significantly alter our conclusions
regarding the importance of inhomogeneities in the UVB spectral shape.

\subsection{The UVB spectral shape}

\begin{figure}
\begin{center}
  \includegraphics[width=0.48\textwidth]{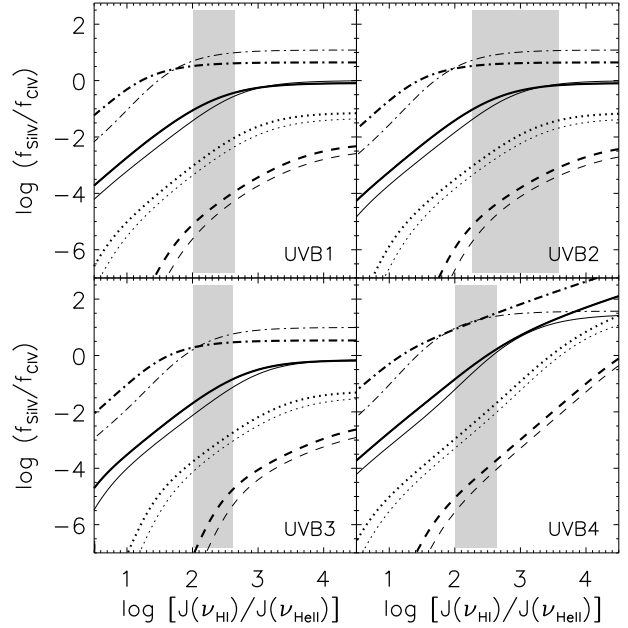}
  \vspace{-0.4cm}
  \caption{The ratio of the \SiIV to \CIV ionisation fractions,
    $f_{\rm SiIV}/f_{\rm CIV}$, for the four UVB models considered in
    this work.  The ratios are plotted as a function of $J(\nu_{\rm
      HI})/J(\nu_{\rm HeII})$, the ratio of specific intensities at
    the \HI and \HeII ionisation edges (see text for details).  In
    each panel, $f_{\rm SiIV}/f_{\rm CIV}$ is shown at four different
    gas densities, $n_{\rm H}=10^{-2}\rm\,cm^{-3}$ (dot-dashed
    curves), $n_{\rm H}=10^{-3}\rm\,cm^{-3}$ (solid curves), $n_{\rm
      H}=10^{-4}\rm\,cm^{-3}$ (dotted curves) and $n_{\rm
      H}=10^{-5}\rm\,cm^{-3}$ (dashed curves).  The thick curves
    assume a gas temperature of $T=32\,500\rm\,K$ while the thin
    curves correspond to $T=65\,000\rm\,K$.  The horizontal extent of
    the shaded regions display the range encompassing 95 per cent of
    all fluctuations around the median $J(\nu_{\rm HI})/J(\nu_{\rm
      HeII})$ for the spatially inhomogeneous UVB models.  {\it Upper
      left:} UVB1.  {\it Upper right:} UVB2.  {\it Lower left:} UVB3.
    {\it Lower right:} UVB4.}
\label{fig:ionfrac}
\end{center}
\end{figure}

\begin{figure}
\begin{center}
  \includegraphics[width=0.48\textwidth]{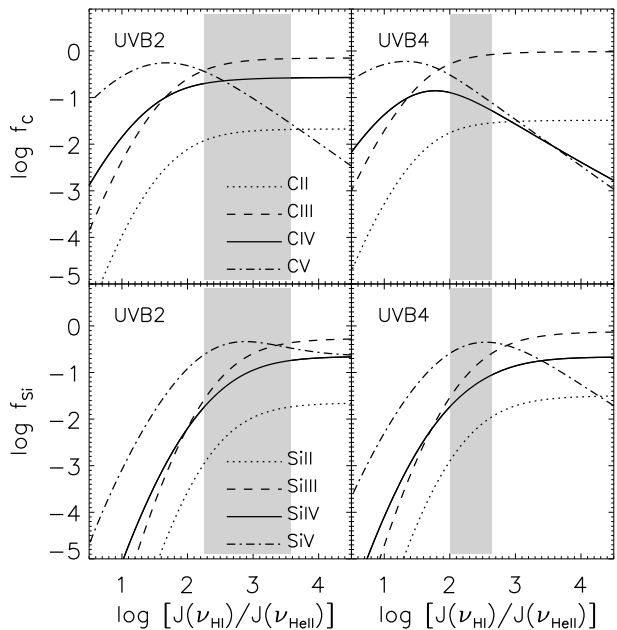}
  \vspace{-0.4cm}
  \caption{Individual ionisation fractions for several different
    ionisation states of carbon (upper panels) and silicon (lower
    panels for model UVB2 (left column) and UVB4 (right column). The
    ionisation fractions are plotted as a function of $J(\nu_{\rm
      HI})/J(\nu_{\rm HeII})$ for gas $n_{\rm H}=10^{-3}\rm\,cm^{-3}$
    and $T=32\,500\rm\,K$. The horizontal extent of the shaded regions
    display the range encompassing 95 per cent of all fluctuations
    around the median $J(\nu_{\rm HI})/J(\nu_{\rm HeII})$ for the
    spatially inhomogeneous UVB models.}
\label{fig:ionfrac2}
\end{center}
\end{figure}

With the typical gas densities and temperatures of the absorption in
hand, we may now examine the effect of varying the spectral shape of
the UVB in more detail.  It is instructive to simply rescale the {\it
  spatially averaged} UVB spectrum for each model by a constant factor
above the \HeII ionisation edge:

\begin{equation}
J(\nu) = \langle J({\bf r},\nu)\rangle \times
\cases{ 1  &($\nu<\nu_{\rm HeII}$),\cr
  \noalign{\vskip3pt} \chi &($\nu \geq \nu_{\rm HeII}$),\cr} \label{eq:rescale}
\end{equation}

\noindent
where $\chi$ is a dimensionless constant.  For model UVB4, this
rescaling also modifies the spectral shape between $E=3-4\rm\,Ryd$.
Smaller (larger) values of $\chi$ thus produce softer (harder) UVB
spectra and a stronger (weaker) break in the spectrum at $4\rm\,Ryd$.

Fig.~\ref{fig:ionfrac} displays the \SiIV to \CIV ionisation fraction,
$f_{\rm SiIV}/f_{\rm CIV}$, for the four UVB models considered in this
work. Ignoring redshift space distortions, this is related to the
optical depth ratio by

\begin{equation} \frac{\tau_{\rm SiIV}}{\tau_{\rm CIV}} \sim \frac{\sigma_{\rm SiIV}}{\sigma_{\rm CIV}}\frac{n_{\rm Si}}{n_{\rm C}}\frac{f_{\rm SiIV}}{f_{\rm CIV}} \simeq 1.7 \frac{f_{\rm SiIV}}{f_{\rm CIV}} 10^{\rm [Si/C] - 0.77}, \label{eq:convert_tau} \end{equation}

\noindent
where $\sigma_{\rm i}$ and $n_{\rm i}$ are the absorption
cross-section and number density for each species.  The ratios are
plotted as a function of the ratio of specific intensities at the \HI
and \HeII ionisation edges, $J(\nu_{\rm HI})/J(\nu_{\rm HeII})$,
following the above rescaling.  The thick curves are computed using
Cloudy for gas with $T=32\,500\rm\,K$, similar to the temperature of
gas responsible for \CIV and \SiIV absorption in our simulated
spectra.  In each panel, the \SiIV to \CIV ratio is shown at four
different gas densities, $n_{\rm H}=10^{-2},\,10^{-3},\,10^{-4}$ and
$10^{-5}\rm \,cm^{-3}$.  These have overdensities at $z=3$ of $\log
\Delta=2.91,\,1.91,\,0.91$ and $-0.09$, respectively.  For comparison,
the median \CIV optical depth weighted overdensity probed by \SiIV and
\CIV absorption with $0.02<\tau_{\rm CIV,SiIV}<5$ for UVB1 is
$\log\Delta_{\tau}=1.7$.

The horizontal extent of the shaded regions in Fig.~\ref{fig:ionfrac}
correspond to the range encompassed by 95 per cent of all fluctuations
from the median $J(\nu_{\rm HI})/J(\nu_{\rm HeII})$ in the {\it
  spatially fluctuating} UVB models.  This provides a rough guide to
the range in $f_{\rm SiIV}/f_{\rm CIV}$ present in the synthetic
spectra as a function of density at fixed temperature.  However, there
are three important points to keep in mind.  Firstly, variations in
the gas temperature\footnote{We do not include the effect of large
  scale ($\ga 50\rm\,Mpc$) spatial fluctuations in the temperature of
  the low density IGM expected during \HeII reionisation
  (\citealt{Theuns02c,McQuinn09}).  However, these fluctuations are
  expected to have a significant impact on large scale, transverse
  correlations in the \Lya forest only; their signatures are much more
  difficult to detect along individual lines-of-sight and on smaller
  scales (\citealt{McQuinn10}).}  will produce additional fluctuations
in $f_{\rm SiIV}/f_{\rm CIV}$ at fixed density.  This is illustrated
by the thin curves in Fig.~\ref{fig:ionfrac}, which are computed for
gas with $T=65\,000\rm\,K$; this temperature lies toward the upper
range of the scatter observed in Fig.~\ref{fig:Trho}. Secondly, only
values of $-1.6\la\log (\tau_{\rm SiIV}/\tau_{\rm CIV})\la 1$ (lower
right panel, Fig.~\ref{fig:ratios}) are readily observable in our
synthetic spectra, which from Eq.~(\ref{eq:convert_tau}) corresponds
to $-1.8\la \log (f_{\rm SiIV}/f_{\rm CIV}) \la 0.8$.  Thirdly, the
range of $f_{\rm SiIV}/f_{\rm CIV}$ fluctuations will in practice be
smaller than indicated by the range of $J(\nu_{\rm HI})/J(\nu_{\rm
  HeII})$ shown by the shading in Fig.~\ref{fig:ionfrac}. Fluctuations
in the UVB spectral shape are {\it maximised} at $4\rm\,Ryd$ where the
mean free path is shortest ({\it e.g.}  Fig.~\ref{fig:UVBspectra}),
whereas \CIV and \SiIV have ionisation potentials above and below
$4\rm\,Ryd$, respectively.  Furthermore, the fluctuations do not have
an equal probability of occurring in this range, and have a higher
probability of lying close to the median ({\it e.g.}
Fig.~\ref{fig:flucPDF}).  The shaded regions thus correspond to an
upper limit in the expected $f_{\rm SiIV}/f_{\rm CIV}$ variation for
each model.

With these points in mind, the $f_{\rm SiIV}/f_{\rm CIV}$ ratio
nevertheless shows only modest variation ($< 1.0$ dex) over the shaded
range in Fig.~\ref{fig:ionfrac} for $n_{\rm H}=10^{-3}\rm\,cm^{-3}$,
with the largest variations occurring for UVB2 and UVB4.  Towards
larger $J(\nu_{\rm HI})/J(\nu_{\rm HeII})$ ({\it i.e.} softer
spectra), $f_{\rm SiIV}/f_{\rm CIV}$ noticeably flattens for all
models, and becomes almost constant for UVB1, UVB2 and UVB3.  Larger
variations in $f_{\rm SiIV}/f_{\rm CIV}$ do occur at lower densities,
especially for model UVB3 where the \SiIV to \CIV ratio changes by
$\sim 2.2$ dex within the shaded range for $n_{\rm
  H}=10^{-5}\rm\,cm^{-3}$.  The \SiIV to \CIV optical depth ratio for
{\it all} pixels in Fig.~\ref{fig:ratios} is slightly broader than the
uniform case for UVB3 at low values as a consequence.  However, from
Eq.~(\ref{eq:convert_tau}) we infer that \CIV and \SiIV absorption
from such low density gas is not detectable in the synthetic spectra.

In Fig.~\ref{fig:ionfrac2} the individual carbon (upper panels) and
silicon (lower panels) ionisation fractions for several different ions
are shown against $J(\nu_{\rm HI})/J(\nu_{\rm HeII})$.  The curves are
again computed assuming densities and temperatures typical of the \CIV
and \SiIV absorption in our simulated spectra, $n_{\rm
  H}=10^{-3}\rm\,cm^{-3}$ and $T=32\,500\rm\,K$.  The left panels
display the results for UVB2, while UVB4 is shown on the right.
Models UVB1 and UVB3 display similar behaviour to UVB2 within the
equivalent ranges for $J(\nu_{\rm HI})/J(\nu_{\rm HeII})$.  Note the
\CIV and \SiIV fractions for model UVB2 are largely insensitive to the
UVB spectral shape for $\log[J(\nu_{\rm HI})/J(\nu_{\rm HeII})]\ga 2$
and $3$, respectively.  This is because the UVB is very soft,
resulting in photo-ionisation rates that are too low to significantly
change the \CIV and \SiIV fractions in the higher density gas
responsible for most of the absorption.  On the other hand, the \CIV
fraction for model UVB4 is more sensitive to an increase in the
strength of the $3-4\rm\,Ryd$ break.  The break at $3\rm\,Ryd$ in this
model produces a larger proportion of carbon in the form of \CIII
relative to \CIV for increasing $J(\nu_{\rm HI})/J(\nu_{\rm HeII})$
(\citealt{Agafonova07,MadauHaardt09,Vasiliev10}).  However, \SiIV
again flattens towards larger values for $J(\nu_{\rm HI})/J(\nu_{\rm
  HeII})$, moderating the change in the \SiIV to \CIV ratio.
Furthermore, as the \CIV fraction drops rapidly at $\log[J(\nu_{\rm
    HI})/J(\nu_{\rm HeII})]>2$ it will become progressively more
difficult to detect the weakening \CIV absorption.

Consequently, as the \HeII ionising photon mean free path becomes
smaller toward higher redshift, the average UVB spectral shape softens
and variations in the {\it observed} \SiIV to \CIV ratio do not become
significantly more pronounced as a result of spatial fluctuations in
the UVB spectral shape.  The \CIV and \SiIV absorption systems
typically originate from overdense regions with $\log\Delta \simeq 2$
in our simulations, where the \SiIV and \CIV fractions change less
rapidly with increasing $J(\nu_{\rm HI})/J(\nu_{\rm HeII})$ compared
to lower density.  It is this behaviour, combined with the smaller
fluctuations expected in the UVB spectral shape at frequencies above
and below the \HeII ionisation edge, which explain the similarity
between the \SiIV to \CIV optical depth ratios for the uniform and
fluctuating models in Fig.~\ref{fig:ratios}.  This is in striking
contrast to the \HeII to \HI ratio, where the same spatial variations
in the UVB spectral shape significantly increase fluctuations in
$\tau_{\rm HeII}/\tau_{\rm HI}$ from lower density gas as the mean
free path is lowered (\citealt{Fardal98,Bolton06,Furlanetto09}).  The
absence of any observational evidence for fluctuations in the
$\tau_{\rm SiIV}/\tau_{\rm CIV}$ ratio (or the inferred silicon to
carbon ratio) is thus not necessarily indicative of a spatially
uniform UVB at $z\simeq 3$.  On the other hand, slightly larger
$\tau_{\rm SiIV}/\tau_{\rm CIV}$ fluctuations are expected at lower
densities, although the weaker absorption from these regions is much
more difficult to detect.  We now examine whether UVB fluctuations can
be detected statistically in the observational data, and whether or
not they contribute to the observed scatter in the IGM metallicity
(\citealt{Rauch97b,Schaye03,Simcoe04}).

\begin{figure*}
\centering
\begin{minipage}{180mm}
\begin{center}
   \psfig{figure=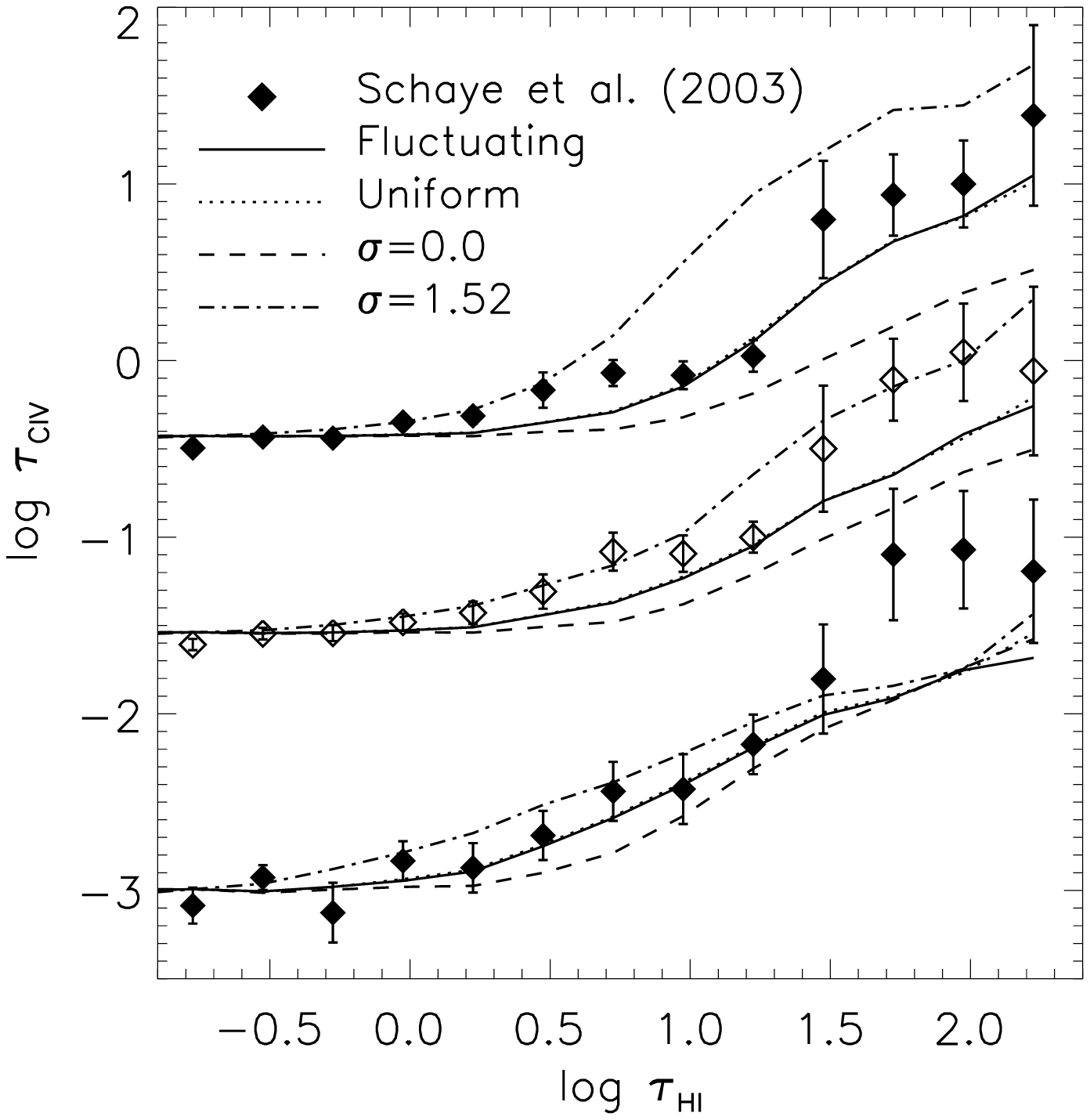,width=0.45\textwidth}
   \psfig{figure=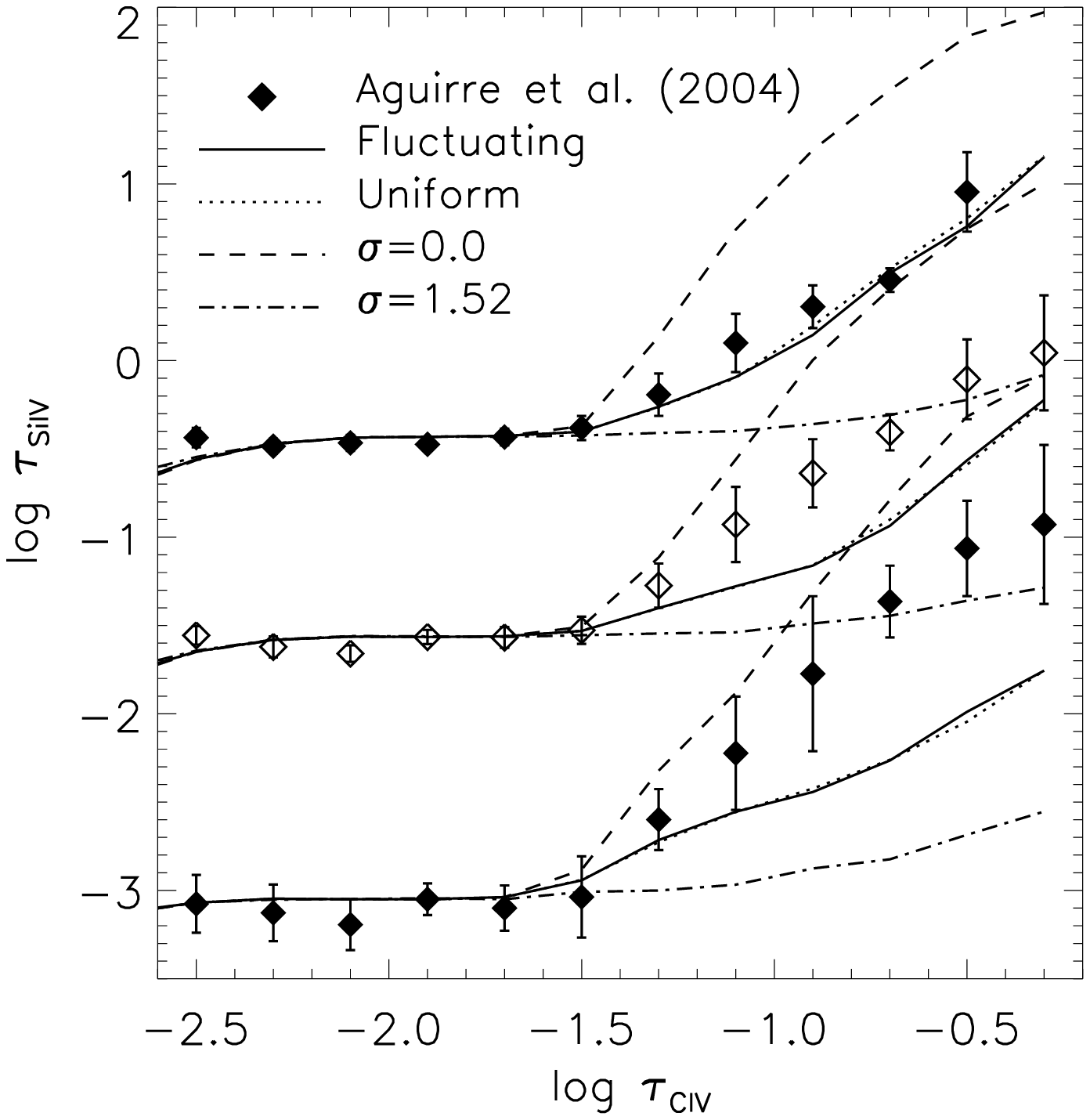,width=0.45\textwidth}
   \vspace{-0.2cm}
  \caption{{\it Left panel:} Comparison of the \CIV pixel optical
    depths as a function of \HI optical depth, obtained from model
    UVB4 to the \citet{Schaye03} observational measurements for
    Q1422$+$230.  From top to bottom, the data points correspond to
    the 84th, 69th and 50th percentiles of the recovered \CIV optical
    depths.  The 84th and 69th percentiles have been offset by $+1.5$
    dex and $+0.75$ dex for clarity.  The solid curves correspond to
    the pixel optical depths recovered using the spatially
    inhomogeneous UVB model, while the dotted curves shows the results
    for the spatially averaged (uniform) spectrum; these two curves
    are almost indistinguishable.  The dashed curves assumes a
    spatially uniform spectrum, but include zero scatter in the
    metallicity at fixed density.  The dot-dashed curves are for
    lognormal scatter drawn from a distribution with $\sigma=1.52$,
    twice the fiducial value of $\sigma=0.76$ . {\it Right panel:} The
    recovered \SiIV optical depth as a function of \CIV optical depth
    for UVB4.  The simulations are compared to the observational data
    of \citet{Aguirre04} for Q1422$+$230. The 84th and 69th
    percentiles are again offset by $+1.5$ dex and $+0.75$ dex for
    clarity, and the different curves are as described for the left
    hand panel.}

\label{fig:pod}

\end{center}
\end{minipage}
\end{figure*}

%%%%%%%%%%%%%%%%%%%%%%%%%%%%%%%%%%%%%%%%%%%%%%%%%%%%%%%%%%%%%%%%%%%%%
%%%%%%%%%%%%%%%%%%%%%%%%%% SECTION 5 %%%%%%%%%%%%%%%%%%%%%%%%%%%%%%%%
%%%%%%%%%%%%%%%%%%%%%%%%%%%%%%%%%%%%%%%%%%%%%%%%%%%%%%%%%%%%%%%%%%%%%

\section{Pixel optical depth analysis} \label{sec:pod}
\subsection{Method}

The pixel optical depth (POD) procedure provides a powerful tool for
statistically analysing metal absorption at low gas densities ({\it
  e.g.}  \citealt{Songaila98,Ellison00,Pieri04}).  Briefly, the
procedure involves recovering optical depths for a base transition
({\it e.g.}  \HI \Lya absorption) on a pixel-by-pixel basis and
pairing these with metal optical depths recovered at the same
redshift.  Given suitable calibration, typically achieved using a
cosmological hydrodynamical simulation, information may then be
derived on the abundance and distribution of metals in the IGM.  The
\CIV POD analysis performed by \cite{Schaye03} found evidence for
scatter in $\rm [C/H]$ at fixed $\tau_{\rm HI}$, described by a normal
distribution in $\rm [C/H]$ with standard deviation $\rm
\sigma([C/H])=0.76-0.23(\log\Delta - 0.5)$ at $z=3$.
\cite{Scannapieco06} and \cite{Pieri06} noted this scatter may be
attributable to spatial variations in the metallicity of the IGM,
although they did not rule out the possibility that spatial
fluctuations in the UVB spectral shape (and hence the assumption of a
uniform ionisation correction) may also play a role.  On the other
hand, \cite{Aguirre04} performed a POD analysis for \SiIV and found no
evidence for additional scatter in the $\tau_{\rm SiIV}/\tau_{\rm
  CIV}$ ratio beyond the aforementioned scatter in the
metallicity. The study concluded that a uniform $\rm [Si/C]$ is
favoured and that inhomogeneities in the spectral shape of the UVB are
small.  However, none of these studies modelled the effect of UVB
fluctuations on the ionisation balance for carbon and silicon in
detail.

We perform a POD analysis on our synthetic spectra using the procedure
outlined by \cite{Aguirre02}.  We consider the optical depths for \HI
\Lya through to \Lyd as well as \CIV and \SiIV absorption.  For the
POD analysis we construct absorption spectra which resemble high
resolution, high signal-to-noise observational data ({\it e.g.}
\citealt{Schaye03}).  The synthetic spectra are convolved with a
Gaussian instrument profile of width $7\rm\,km\,s^{-1}$, resampled
onto pixels of width $3.1 \rm \,km\,s^{-1}$ and Gaussian distributed
noise with $S/N=100$ is added.  Continuum fitting errors are mimicked
by performing an iterative continuum correction to the synthetic
spectra.  We compute the median transmitted flux in each synthetic
line-of-sight and deselect all pixels below $1\sigma$ of this value,
where $\sigma$ is the rms noise amplitude in each pixel.  The median
flux is computed again for the remaining pixels, and the procedure is
then continued until convergence is reached.

\subsection{Comparison to observational data}

The results of our POD analysis are presented in Fig.~\ref{fig:pod}.
The data points with 1$\sigma$ error bars in the left panel correspond
to the \citet{Schaye03} observational measurements for
Q1422$+$230. These data have a median absorption redshift of
$z=3.225$, at slightly higher redshift than our synthetic absorption
spectra.  From top to bottom, the three sets of data points correspond
to the 84th, 69th and 50th percentiles of the recovered \CIV optical
depths; the 84th and 69th percentiles have been offset by $+1.5$ dex
and $+0.75$ dex for clarity.  The curves display the recovered \CIV
pixel optical depth against \HI optical depth for spectra constructed
using model UVB4, which we find has the largest impact on the
recovered optical depths.  The solid curves are obtained from the
spatially inhomogeneous UVB model, while the dotted curves corresponds
to the optical depths recovered from the spatially averaged spectrum.
The curves are almost indistinguishable, indicating that the predicted
spatial inhomogeneities in the UVB spectral shape approaching \HeII
reionisation will not significantly impact on the ionisation
correction for C$\,\rm \scriptstyle IV$.

The dashed and dot-dashed curves in Fig.~\ref{fig:pod} use the
spatially uniform UVB model, but now also include different
assumptions for the lognormal scatter in the IGM metallicity at fixed
density.  The dashed curves assume zero scatter, $\sigma=0$, while the
dot-dashed curves correspond to $\sigma=1.52$.  The impact on the
recovered optical depths is especially prominent for the higher
percentiles, suggesting that spatial variations in metallicity rather
than the UVB spectral shape will dominate any scatter.  As noted by
\cite{Schaye03}, the fiducial model with $\sigma=0.76$ is in somewhat
better agreement with the higher percentiles.

In the right hand panel we compare the same model to the POD
measurements for \SiIV and \CIV presented by \cite{Aguirre04} for the
same quasar.  Again, the recovered optical depths are very similar for
the spatially inhomogeneous and uniform UVB models, and scatter in the
metallicity has a much larger impact on the recovered optical depths.
The fiducial model with $\sigma=0.76$ is again in better agreement
with the higher percentiles, although the agreement is poorer for the
median.  Note also, in contrast to the \CIV and \HI pixel optical
depths, lognormal scatter in the metallicity at fixed density {\it
  lowers} each percentile for $\tau_{\rm SiIV}$ as a function of
$\tau_{\rm CIV}$.  The explanation for this behaviour is that scatter
in metallicity is added to {\it both} of the pixel optical depths, in
contrast to the left hand panel of Fig.~\ref{fig:pod}.  Adding this
scatter enables the detection of both \CIV and \SiIV absorption in
more pixels at lower densities due to their higher metallicities;
these pixels are otherwise hidden in the flat (noise and continuum
error dominated) part of the correlation at $\log \tau_{\rm
  CIV}<-1.5$.  However, for this UVB model the \SiIV fraction (and
hence optical depth) decreases more rapidly with decreasing gas
density relative to the \CIV fraction.  Consequently, a greater
fraction of \CIV optical depths with low $\tau_{\rm SiIV}$ are now
{\it detectable}, leading to the lowering of the $\tau_{\rm SiIV}$
percentiles observed in Fig.~\ref{fig:pod}.

Finally, we caution the reader not to take the differences between the
observational data and simulations too seriously; the \cite{Schaye03}
and \cite{Aguirre04} metallicities we have used to construct our
spectra are derived under the assumption of a different UVB model.
There is therefore no reason to expect our model to match the data
exactly using these metallicities; the comparison here is instead
largely illustrative and one may always fine tune the metallicity to
enable a given UVB model to match the data.  The key point is that we
clearly expect variations in the {\it spatial distribution} of metals
to dominate any scatter in the \CIV and \SiIV optical depths at fixed
density, even in the presence of the spatially inhomogeneous UVB
expected at the tail-end of \HeII reionisation.

%%%%%%%%%%%%%%%%%%%%%%%%%%%%%%%%%%%%%%%%%%%%%%%%%%%%%%%%%%%%%%%%%%%%%	
%%%%%%%%%%%%%%%%%%%%%%%%%% SECTION 6 %%%%%%%%%%%%%%%%%%%%%%%%%%%%%%%%
%%%%%%%%%%%%%%%%%%%%%%%%%%%%%%%%%%%%%%%%%%%%%%%%%%%%%%%%%%%%%%%%%%%%%

\section{Conclusions}

We use a large hydrodynamical simulation of the IGM combined with a
toy model for spatial inhomogeneities in the UVB spectral shape to
investigate the impact of spectral hardness fluctuations on the
ionisation balance of intergalactic carbon and silicon at $z\simeq 3$.
We construct synthetic quasar absorption spectra from the simulations,
assuming the metallicity of the IGM traces the underlying gas density
(\citealt{Schaye03,Aguirre04}).  We carefully examine the impact of
the UVB fluctuations on the \SiIV and \CIV optical depths.  Four
different spatially inhomogeneous UVB models which employ a variety of
different assumptions for the UVB spectral shape are considered in our
analysis.  We reconfirm that fluctuations in the UVB spectral shape
expected at the tail-end of \HeII reionisation have a significant
impact on the \HeII to \HI optical depth ratio (see also
\citealt{Fardal98,Bolton06,Furlanetto09}). However, some of the lowest
values for this ratio also result from lines which are predominantly
thermally broadened (\citealt{Fechner07}).  On the other hand, while
the \SiIV to \CIV ratio is indeed sensitive to the average spectral
shape of the UVB, we find the predicted fluctuations have little
impact on $\tau_{\rm SiIV}/\tau_{\rm CIV}$ measured in our synthetic
spectra.

The majority of the detectable \SiIV and \CIV absorption in our
synthetic spectra originates from regions with $\log \Delta \simeq
1.5-2 $ and $T\simeq 35\,000\rm\,K$.  These absorbers are
predominantly photo-ionised in our simulations, and we may thus
exclude collisional ionisation as a possible explanation for the small
impact of UVB fluctuations on our simulated spectra.  Instead, we find
the ratio of observable \SiIV to \CIV optical depths varies relatively
little considering the wide range of fluctuations in the UVB spectral
shape.  This is in part because of the longer mean free path for
photons above and below the \HeII ionisation edge, which results in
smaller fluctuations in the UVB spectral shape at these frequencies.
However, as fluctuations in the UVB spectral shape become larger as
the \HeII opacity increases, the spatially averaged UVB spectral shape
becomes softer; UVB models which produce the largest fluctuations in
$\tau_{\rm HeII}/\tau_{\rm HI}$ also have photo-ionisation rates which
are too low to have a significant impact on the observed \SiIV to \CIV
ratio.  At lower gas densities, or for UVB models which predict a
larger fraction of \CIII relative to \CIV ({\it e.g.} for \HeII Lyman
series absorption), the expected variation in the \SiIV to \CIV ratio
can be slightly larger, but it is more difficult to detect due to the
correspondingly smaller gas densities and/or fraction of triply
ionised carbon produced.

Finally, we briefly examine the observational consequences for studies
of the IGM metallicity using \CIV and \SiIV absorption at $z\simeq 3$.
We perform a pixel optical depth analysis on our synthetic spectra,
and find that the predicted UVB hardness fluctuations will have little
impact on observations compared to spatial variations in the IGM
metallicity.  We conclude that the lack of any observed fluctuations
in the $\tau_{\rm SiIV}/\tau_{\rm CIV}$ ratio does not provide a
stringent limit on the non-uniformity of the UVB spectral shape, and
in particular does not preclude the possibility of \HeII reionisation
completing around $z\simeq 2-3$.  On the other hand, we confirm the
observed scatter in the IGM metallicity inferred from \CIV and \SiIV
absorption $z\simeq 2-3$ is likely to be intrinsic, reinforcing its
potential as a powerful constraint on intergalactic metal enrichment
scenarios.

\section*{Acknowledgments}

The authors thank the staff of the Institute of Astronomy and the
Kavli Institute for Cosmology, Cambridge, for their hospitality during
the completion of this work, and acknowledge valuable conversations
with George Becker, Benedetta Ciardi, Luca Graziani, Martin Haehnelt,
Matthew Pieri, Michael Rauch and Stuart Wyithe.  We also thank Rob
Simcoe for drawing our attention to his latest results ahead of
publication, and the anonymous referee for a constructive report which
helped improve this paper.  The hydrodynamical simulation used in this
work was performed using the Darwin Supercomputer of the University of
Cambridge High Performance Computing Service
(http://www.hpc.cam.ac.uk/), provided by Dell Inc. using Strategic
Research Infrastructure Funding from the Higher Education Funding
Council for England.  JSB acknowledges the support of an ARC
Australian postdoctoral fellowship (DP0984947).  MV is partly
supported by: ASI/AAE theory grant, INFN-PD51 grant, PRIN-MIUR,
PRIN-INAF and by the FP7 ERC starting grant ``CosmoIGM''.


\begin{thebibliography}{99}


\bibitem[{{Agafonova} {et~al.}(2005){Agafonova}, {Centuri{\'o}n}, {Levshakov},
  \& {Molaro}}]{Agafonova05}
{Agafonova}, I.~I., {Centuri{\'o}n}, M., {Levshakov}, S.~A., \& {Molaro}, P.
  2005, \aap, 441, 9

\bibitem[{{Agafonova} {et~al.}(2007){Agafonova}, {Levshakov}, {Reimers},
  {Fechner}, {Tytler}, {Simcoe}, \& {Songaila}}]{Agafonova07}
{Agafonova}, I.~I., {Levshakov}, S.~A., {Reimers}, D., {Fechner}, C., {Tytler},
  D., {Simcoe}, R.~A., \& {Songaila}, A. 2007, \aap, 461, 893

\bibitem[{{Aguirre} {et~al.}(2008){Aguirre}, {Dow-Hygelund}, {Schaye}, \&
  {Theuns}}]{Aguirre08}
{Aguirre}, A., {Dow-Hygelund}, C., {Schaye}, J., \& {Theuns}, T. 2008, \apj,
  689, 851

\bibitem[{{Aguirre} {et~al.}(2004){Aguirre}, {Schaye}, {Kim}, {Theuns},
  {Rauch}, \& {Sargent}}]{Aguirre04}
{Aguirre}, A., {Schaye}, J., {Kim}, T., {Theuns}, T., {Rauch}, M., \&
  {Sargent}, W.~L.~W. 2004, \apj, 602, 38

\bibitem[{{Aguirre} {et~al.}(2002){Aguirre}, {Schaye}, \& {Theuns}}]{Aguirre02}
{Aguirre}, A., {Schaye}, J., \& {Theuns}, T. 2002, \apj, 576, 1

\bibitem[{{Aracil} {et~al.}(2004){Aracil}, {Petitjean}, {Pichon}, \&
  {Bergeron}}]{Aracil04}
{Aracil}, B., {Petitjean}, P., {Pichon}, C., \& {Bergeron}, J. 2004, \aap, 419,
  811

\bibitem[{{Asplund} {et~al.}(2009){Asplund}, {Grevesse}, {Sauval}, \&
  {Scott}}]{Asplund09}
{Asplund}, M., {Grevesse}, N., {Sauval}, A.~J., \& {Scott}, P. 2009, \araa, 47,
  481

\bibitem[{{Becker} {et~al.}(2011){Becker}, {Bolton}, {Haehnelt}, \&
  {Sargent}}]{Becker10}
{Becker}, G.~D., {Bolton}, J.~S., {Haehnelt}, M.~G., \& {Sargent}, W.~L.~W.
  2011, MNRAS, 410, 1096

\bibitem[{{Bianchi} {et~al.}(2001){Bianchi}, {Cristiani}, \& {Kim}}]{Bianchi01}
{Bianchi}, S., {Cristiani}, S., \& {Kim}, T.-S. 2001, \aap, 376, 1

\bibitem[{{Boksenberg} {et~al.}(2003){Boksenberg}, {Sargent}, \&
    {Rauch}}]{Boksenberg03} {Boksenberg}, A., {Sargent}, W.~L.~W., \&
  {Rauch}, M. 2003, preprint, astro-ph/0307557

\bibitem[{{Bolton} {et~al.}(2006){Bolton}, {Haehnelt}, {Viel}, \&
  {Carswell}}]{Bolton06}
{Bolton}, J.~S., {Haehnelt}, M.~G., {Viel}, M., \& {Carswell}, R.~F. 2006,
  \mnras, 366, 1378

\bibitem[{{Bolton} {et~al.}(2005){Bolton}, {Haehnelt}, {Viel}, \&
  {Springel}}]{Bolton05}
{Bolton}, J.~S., {Haehnelt}, M.~G., {Viel}, M., \& {Springel}, V. 2005, \mnras,
  357, 1178

\bibitem[{{Cen} \& {Chisari}(2010)}]{CenChisari10}
{Cen}, R. \& {Chisari}, N.~E. 2010, ApJ submitted, arXiv:1005.1451

\bibitem[{{Cen} {et~al.}(2005){Cen}, {Nagamine}, \& {Ostriker}}]{Cen05}
{Cen}, R., {Nagamine}, K., \& {Ostriker}, J.~P. 2005, \apj, 635, 86

\bibitem[{{Cowie} {et~al.}(1995){Cowie}, {Songaila}, {Kim}, \& {Hu}}]{Cowie95}
{Cowie}, L.~L., {Songaila}, A., {Kim}, T.-S., \& {Hu}, E.~M. 1995, \aj, 109,
  1522

\bibitem[{{Croft}(2004)}]{Croft04}
{Croft}, R.~A.~C. 2004, \apj, 610, 642

\bibitem[{{Croft} {et~al.}(1997){Croft}, {Weinberg}, {Katz}, \&
  {Hernquist}}]{Croft97}
{Croft}, R.~A.~C., {Weinberg}, D.~H., {Katz}, N., \& {Hernquist}, L. 1997,
  \apj, 488, 532

\bibitem[{{Dixon} \& {Furlanetto}(2009)}]{DixonFurlanetto09}
{Dixon}, K.~L. \& {Furlanetto}, S.~R. 2009, \apj, 706, 970

\bibitem[{{D'Odorico} {et~al.}(2010){D'Odorico}, {Calura}, {Cristiani}, \&
  {Viel}}]{Dodorico10}
{D'Odorico}, V., {Calura}, F., {Cristiani}, S., \& {Viel}, M. 2010, \mnras,
  401, 2715

\bibitem[{{Ellison} {et~al.}(2000){Ellison}, {Songaila}, {Schaye}, \&
  {Pettini}}]{Ellison00}
{Ellison}, S.~L., {Songaila}, A., {Schaye}, J., \& {Pettini}, M. 2000, \aj,
  120, 1175

\bibitem[{{Fardal} {et~al.}(1998){Fardal}, {Giroux}, \& {Shull}}]{Fardal98}
{Fardal}, M.~A., {Giroux}, M.~L., \& {Shull}, J.~M. 1998, \aj, 115, 2206

\bibitem[{{Faucher-Gigu{\`e}re} {et~al.}(2009){Faucher-Gigu{\`e}re}, {Lidz},
  {Zaldarriaga}, \& {Hernquist}}]{Faucher09}
{Faucher-Gigu{\`e}re}, C., {Lidz}, A., {Zaldarriaga}, M., \& {Hernquist}, L.
  2009, \apj, 703, 1416

\bibitem[{{Faucher-Gigu{\`e}re} {et~al.}(2008){Faucher-Gigu{\`e}re}, {Lidz},
  {Hernquist}, \& {Zaldarriaga}}]{Faucher08b}
{Faucher-Gigu{\`e}re}, C.-A., {Lidz}, A., {Hernquist}, L., \& {Zaldarriaga}, M.
  2008, \apj, 688, 85

\bibitem[{{Fechner} \& {Reimers}(2007)}]{Fechner07}
{Fechner}, C. \& {Reimers}, D. 2007, \aap, 463, 69

\bibitem[{{Fechner} {et~al.}(2006){Fechner}, {Reimers}, {Kriss}, {Baade},
  {Blair}, {Giroux}, {Green}, {Moos}, {Morton}, {Scott}, {Shull}, {Simcoe},
  {Songaila}, \& {Zheng}}]{Fechner06}
{Fechner}, C. et al. 2006, \aap, 455, 91

\bibitem[{{Fechner} \& {Richter}(2009)}]{FechnerRichter09}
{Fechner}, C. \& {Richter}, P. 2009, \aap, 496, 31

\bibitem[{{Ferland} {et~al.}(1998){Ferland}, {Korista}, {Verner}, {Ferguson},
  {Kingdon}, \& {Verner}}]{Ferland98}
{Ferland}, G.~J., {Korista}, K.~T., {Verner}, D.~A., {Ferguson}, J.~W.,
  {Kingdon}, J.~B., \& {Verner}, E.~M. 1998, \pasp, 110, 761

\bibitem[{{Fox} {et~al.}(2008){Fox}, {Bergeron}, \& {Petitjean}}]{Fox08}
{Fox}, A.~J., {Bergeron}, J., \& {Petitjean}, P. 2008, \mnras, 388, 1557

\bibitem[{{Furlanetto} \& {Lidz}(2010)}]{FurlanettoLidz10}
{Furlanetto}, S. \& {Lidz}, A. 2010, ApJ submitted, arXiv:1008.4609

\bibitem[{{Furlanetto}(2009{\natexlab{a}})}]{Furlanetto09}
{Furlanetto}, S.~R. 2009{\natexlab{a}}, \apj, 703, 702

\bibitem[{{Furlanetto}(2009{\natexlab{b}})}]{Furlanetto09b}
{Furlanetto}, S.~R. 2009{\natexlab{b}}, \apj, 700, 1666

\bibitem[{{Furlanetto} \& {Oh}(2008)}]{FurlanettoOh08}
{Furlanetto}, S.~R. \& {Oh}, S.~P. 2008, \apj, 682, 14

\bibitem[{{Giroux} \& {Shull}(1997)}]{GirouxShull97}
{Giroux}, M.~L. \& {Shull}, J.~M. 1997, \aj, 113, 1505

\bibitem[{{Haardt} \& {Madau}(1996)}]{HaardtMadau96}
{Haardt}, F. \& {Madau}, P. 1996, \apj, 461, 20

\bibitem[{{Haardt} \& {Madau}(2001)}]{HaardtMadau01}
{Haardt}, F. \& {Madau}, P. 2001, in Clusters of Galaxies and the High Redshift
  Universe Observed in X-rays, {Neumann}, D.~M. \& {Tran}, J.~T.~V. ed.,
  astro-ph/0106018

\bibitem[{{Haehnelt} {et~al.}(2001){Haehnelt}, {Madau}, {Kudritzki}, \&
  {Haardt}}]{Haehnelt01}
{Haehnelt}, M.~G., {Madau}, P., {Kudritzki}, R., \& {Haardt}, F. 2001, \apjl,
  549, L151

\bibitem[{{Hopkins} {et~al.}(2007){Hopkins}, {Richards}, \&
  {Hernquist}}]{Hopkins07}
{Hopkins}, P.~F., {Richards}, G.~T., \& {Hernquist}, L. 2007, \apj, 654, 731

\bibitem[{{Kim} {et~al.}(2002){Kim}, {Carswell}, {Cristiani}, {D'Odorico}, \&
  {Giallongo}}]{Kim02}
{Kim}, T.-S., {Carswell}, R.~F., {Cristiani}, S., {D'Odorico}, S., \&
  {Giallongo}, E. 2002, \mnras, 335, 555

\bibitem[{{Kirkman} {et~al.}(2005){Kirkman}, {Tytler}, {Suzuki}, {Melis},
  {Hollywood}, {James}, {So}, {Lubin}, {Jena}, {Norman}, \&
  {Paschos}}]{Kirkman05}
{Kirkman}, D. et al. 2005, \mnras, 360, 1373

\bibitem[{{Komatsu} {et~al.}(2009){Komatsu}, {Dunkley}, {Nolta}, {Bennett},
  {Gold}, {Hinshaw}, {Jarosik}, {Larson}, {Limon}, {Page}, {Spergel},
  {Halpern}, {Hill}, {Kogut}, {Meyer}, {Tucker}, {Weiland}, {Wollack}, \&
  {Wright}}]{Komatsu09}
{Komatsu}, E. et al. 2009, \apjs, 180, 330

\bibitem[{{Madau} \& {Haardt}(2009)}]{MadauHaardt09}
{Madau}, P. \& {Haardt}, F. 2009, \apjl, 693, L100

\bibitem[{{Madau} {et~al.}(1999){Madau}, {Haardt}, \& {Rees}}]{Madau99}
{Madau}, P., {Haardt}, F., \& {Rees}, M.~J. 1999, \apj, 514, 648

\bibitem[{{Martin} {et~al.}(2010){Martin}, {Scannapieco}, {Ellison}, {Hennawi},
  {Djorgovski}, \& {Fournier}}]{Martin10}
{Martin}, C.~L., {Scannapieco}, E., {Ellison}, S.~L., {Hennawi}, J.~F.,
  {Djorgovski}, S.~G., \& {Fournier}, A.~P. 2010, \apj, 721, 174

\bibitem[{{Maselli} \& {Ferrara}(2005)}]{Maselli05}
{Maselli}, A. \& {Ferrara}, A. 2005, \mnras, 364, 1429

\bibitem[{{McQuinn}(2009)}]{McQuinn09b}
{McQuinn}, M. 2009, \apjl, 704, L89

\bibitem[{{McQuinn} {et~al.}(2010){McQuinn}, {Hernquist}, {Lidz}, \&
  {Zaldarriaga}}]{McQuinn10}
{McQuinn}, M., {Hernquist}, L., {Lidz}, A., \& {Zaldarriaga}, M. 2010, MNRAS submitted, arXiv:1010.5250
 

\bibitem[{{McQuinn} {et~al.}(2009){McQuinn}, {Lidz}, {Zaldarriaga},
  {Hernquist}, {Hopkins}, {Dutta}, \& {Faucher-Gigu{\`e}re}}]{McQuinn09}
{McQuinn}, M., {Lidz}, A., {Zaldarriaga}, M., {Hernquist}, L., {Hopkins},
  P.~F., {Dutta}, S., \& {Faucher-Gigu{\`e}re}, C.-A. 2009, \apj, 694, 842

\bibitem[{{Meiksin} \& {White}(2004)}]{MeiksinWhite04}
{Meiksin}, A. \& {White}, M. 2004, \mnras, 350, 1107

\bibitem[{{Miralda-Escud{\' e}} {et~al.}(2000){Miralda-Escud{\' e}},
  {Haehnelt}, \& {Rees}}]{MiraldaEscude00}
{Miralda-Escud{\' e}}, J., {Haehnelt}, M., \& {Rees}, M.~J. 2000, \apj, 530, 1

\bibitem[{{Miralda-Escud{\'e}}(2005)}]{Miralda05}
{Miralda-Escud{\'e}}, J. 2005, \apjl, 620, L91

\bibitem[{{Morton}(2003)}]{Morton03}
{Morton}, D.~C. 2003, \apjs, 149, 205

\bibitem[{{Muzahid} {et~al.}(2010){Muzahid}, {Srianand}, \&
  {Petitjean}}]{Muzahid10}
{Muzahid}, S., {Srianand}, R., \& {Petitjean}, P. 2010, MNRAS in press, arXiv:1008.4132

\bibitem[{{Oppenheimer} \& {Dav{\'e}}(2006)}]{Oppenheimer06}
{Oppenheimer}, B.~D. \& {Dav{\'e}}, R. 2006, \mnras, 373, 1265

\bibitem[{{Oppenheimer} {et~al.}(2009){Oppenheimer}, {Dav{\'e}}, \&
  {Finlator}}]{Oppenheimer09}
{Oppenheimer}, B.~D., {Dav{\'e}}, R., \& {Finlator}, K. 2009, \mnras, 396, 729

\bibitem[{{Paschos} {et~al.}(2007){Paschos}, {Norman}, {Bordner}, \&
  {Harkness}}]{Paschos07}
{Paschos}, P., {Norman}, M.~L., {Bordner}, J.~O., \& {Harkness}, R. 2007, preprint, arXiv:0711.1904

\bibitem[{{Petitjean} {et~al.}(1993){Petitjean}, {Webb}, {Rauch}, {Carswell},
  \& {Lanzetta}}]{Petitjean93}
{Petitjean}, P., {Webb}, J.~K., {Rauch}, M., {Carswell}, R.~F., \& {Lanzetta},
  K. 1993, \mnras, 262, 499

\bibitem[{{Pieri} \& {Haehnelt}(2004)}]{Pieri04}
{Pieri}, M.~M. \& {Haehnelt}, M.~G. 2004, \mnras, 347, 985

\bibitem[{{Pieri} {et~al.}(2006){Pieri}, {Schaye}, \& {Aguirre}}]{Pieri06}
{Pieri}, M.~M., {Schaye}, J., \& {Aguirre}, A. 2006, \apj, 638, 45

\bibitem[{{Rauch} {et~al.}(1997){Rauch}, {Haehnelt}, \&
  {Steinmetz}}]{Rauch97b}
{Rauch}, M., {Haehnelt}, M.~G., \& {Steinmetz}, M. 1997, \apj,
  481, 601


\bibitem[{{Reichardt} {et~al.}(2009){Reichardt}, {Ade}, {Bock}, {Bond},
  {Brevik}, {Contaldi}, {Daub}, {Dempsey}, {Goldstein}, {Holzapfel}, {Kuo},
  {Lange}, {Lueker}, {Newcomb}, {Peterson}, {Ruhl}, {Runyan}, \&
  {Staniszewski}}]{Reichardt09}
{Reichardt}, C.~L. et al. 2009, \apj, 694, 1200


\bibitem[{{Savaglio} {et~al.}(1997){Savaglio}, {Cristiani}, {D'Odorico},
  {Fontana}, {Giallongo}, \& {Molaro}}]{Savaglio97}
{Savaglio}, S., {Cristiani}, S., {D'Odorico}, S., {Fontana}, A., {Giallongo},
  E., \& {Molaro}, P. 1997, \aap, 318, 347

\bibitem[{{Scannapieco} {et~al.}(2006){Scannapieco}, {Pichon}, {Aracil},
  {Petitjean}, {Thacker}, {Pogosyan}, {Bergeron}, \&
  {Couchman}}]{Scannapieco06}
{Scannapieco}, E., {Pichon}, C., {Aracil}, B., {Petitjean}, P., {Thacker},
  R.~J., {Pogosyan}, D., {Bergeron}, J., \& {Couchman}, H.~M.~P. 2006, \mnras,
  365, 615

\bibitem[{{Schaye}(2006)}]{Schaye06}
{Schaye}, J. 2006, \apj, 643, 59

\bibitem[{{Schaye} {et~al.}(2003){Schaye}, {Aguirre}, {Kim}, {Theuns}, {Rauch},
  \& {Sargent}}]{Schaye03}
{Schaye}, J., {Aguirre}, A., {Kim}, T., {Theuns}, T., {Rauch}, M., \&
  {Sargent}, W.~L.~W. 2003, \apj, 596, 768

\bibitem[{{Schaye} {et~al.}(2007){Schaye}, {Carswell}, \& {Kim}}]{Schaye07}
{Schaye}, J., {Carswell}, R.~F., \& {Kim}, T. 2007, \mnras, 379, 1169

\bibitem[{{Schaye} {et~al.}(2000){Schaye}, {Rauch}, {Sargent}, \&
  {Kim}}]{Schaye00m}
{Schaye}, J., {Rauch}, M., {Sargent}, W.~L.~W., \& {Kim}, T.-S. 2000, \apjl,
  541, L1

\bibitem[{{Scott} {et~al.}(2004){Scott}, {Kriss}, {Brotherton}, {Green},
  {Hutchings}, {Shull}, \& {Zheng}}]{Scott04}
{Scott}, J.~E., {Kriss}, G.~A., {Brotherton}, M., {Green}, R.~F., {Hutchings},
  J., {Shull}, J.~M., \& {Zheng}, W. 2004, \apj, 615, 135

\bibitem[{{Shen} {et~al.}(2010){Shen}, {Wadsley}, \& {Stinson}}]{Shen10}
{Shen}, S., {Wadsley}, J., \& {Stinson}, G. 2010, \mnras, 407, 1581

\bibitem[{{Shull} {et~al.}(2004){Shull}, {Tumlinson}, {Giroux}, {Kriss}, \&
  {Reimers}}]{Shull04}
{Shull}, J.~M., {Tumlinson}, J., {Giroux}, M.~L., {Kriss}, G.~A., \& {Reimers},
  D. 2004, \apj, 600, 570

\bibitem[{{Shull} {et~al.}(2010){Shull}, {France}, {Danforth}, {Smith}, \&
  {Tumlinson}}]{Shull10}
{Shull}, M., {France}, K., {Danforth}, C., {Smith}, B., \& {Tumlinson}, J.
  2010, \apj, 722, 1312

\bibitem[{{Simcoe} {et~al.}(2004){Simcoe}, {Sargent}, \& {Rauch}}]{Simcoe04}
{Simcoe}, R.~A., {Sargent}, W.~L.~W., \& {Rauch}, M. 2004, \apj, 606, 92

\bibitem[{{Songaila}(1998)}]{Songaila98}
{Songaila}, A. 1998, \aj, 115, 2184

\bibitem[{{Songaila}(2005)}]{Songaila05}
{Songaila}, A. 2005, \aj, 130, 1996

\bibitem[{{Songaila} \& {Cowie}(1996)}]{Songaila96}
{Songaila}, A. \& {Cowie}, L.~L. 1996, \aj, 112, 335

\bibitem[{{Songaila} {et~al.}(1995){Songaila}, {Hu}, \& {Cowie}}]{Songaila95}
{Songaila}, A., {Hu}, E.~M., \& {Cowie}, L.~L. 1995, \nat, 375, 124

\bibitem[{{Springel}(2005)}]{Springel05}
{Springel}, V. 2005, \mnras, 364, 1105

\bibitem[{{Telfer} {et~al.}(2002){Telfer}, {Zheng}, {Kriss}, \&
  {Davidsen}}]{Telfer02}
{Telfer}, R.~C., {Zheng}, W., {Kriss}, G.~A., \& {Davidsen}, A.~F. 2002, \apj,
  565, 773

\bibitem[{{Tescari} {et~al.}(2010){Tescari}, {Viel}, {D'Odorico}, {Cristiani},
  {Calura}, {Borgani}, \& {Tornatore}}]{Tescari10}
{Tescari}, E., {Viel}, M., {D'Odorico}, V., {Cristiani}, S., {Calura}, F.,
  {Borgani}, S., \& {Tornatore}, L. 2010, MNRAS submitted, arXiv:1007.1628

\bibitem[{{Theuns} {et~al.}(1998){Theuns}, {Leonard}, {Efstathiou}, {Pearce},
  \& {Thomas}}]{Theuns98}
{Theuns}, T., {Leonard}, A., {Efstathiou}, G., {Pearce}, F.~R., \& {Thomas},
  P.~A. 1998, \mnras, 301, 478

\bibitem[{{Theuns} {et~al.}(2002{\natexlab{a}}){Theuns}, {Viel}, {Kay},
  {Schaye}, {Carswell}, \& {Tzanavaris}}]{Theuns02b}
{Theuns}, T., {Viel}, M., {Kay}, S., {Schaye}, J., {Carswell}, R.~F., \&
  {Tzanavaris}, P. 2002{\natexlab{a}}, \apjl, 578, L5

\bibitem[{{Theuns} {et~al.}(2002{\natexlab{b}}){Theuns}, {Zaroubi}, {Kim},
  {Tzanavaris}, \& {Carswell}}]{Theuns02c}
{Theuns}, T., {Zaroubi}, S., {Kim}, T.-S., {Tzanavaris}, P., \& {Carswell},
  R.~F. 2002{\natexlab{b}}, \mnras, 332, 367

\bibitem[{{Tittley} \& {Meiksin}(2007)}]{Tittley07}
{Tittley}, E.~R. \& {Meiksin}, A. 2007, \mnras, 380, 1369

\bibitem[{{Tytler} {et~al.}(2009){Tytler}, {Gleed}, {Melis}, {Chapman},
  {Kirkman}, {Lubin}, {Paschos}, {Jena}, \& {Crotts}}]{Tytler09b}
{Tytler}, D. et al. 2009, \mnras, 392, 1539

\bibitem[{{Vasiliev} {et~al.}(2010){Vasiliev}, {Sethi}, \& {Nath}}]{Vasiliev10}
{Vasiliev}, E.~O., {Sethi}, S.~K., \& {Nath}, B.~B. 2010, \apj, 719, 1343

\bibitem[{{Vladilo} {et~al.}(2003){Vladilo}, {Centuri{\'o}n}, {D'Odorico}, \&
  {P{\'e}roux}}]{Vladilo03}
{Vladilo}, G., {Centuri{\'o}n}, M., {D'Odorico}, V., \& {P{\'e}roux}, C. 2003,
  \aap, 402, 487

\bibitem[{{Wiersma} {et~al.}(2010){Wiersma}, {Schaye}, {Dalla Vecchia},
  {Booth}, {Theuns}, \& {Aguirre}}]{Wiersma10}
{Wiersma}, R.~P.~C., {Schaye}, J., {Dalla Vecchia}, C., {Booth}, C.~M.,
  {Theuns}, T., \& {Aguirre}, A. 2010, MNRAS, 409, 132

\bibitem[{{Wild} {et~al.}(2008){Wild}, {Kauffmann}, {White}, {York}, {Lehnert},
  {Heckman}, {Hall}, {Khare}, {Lundgren}, {Schneider}, \& {vanden
  Berk}}]{Wild08}
{Wild}, V. et al. 2008, \mnras, 388, 227

\bibitem[{{Worseck} {et~al.}(2007){Worseck}, {Fechner}, {Wisotzki}, \&
  {Dall'Aglio}}]{Worseck07}
{Worseck}, G., {Fechner}, C., {Wisotzki}, L., \& {Dall'Aglio}, A. 2007, \aap,
  473, 805

\bibitem[{{Zheng} {et~al.}(2004){Zheng}, {Kriss}, {Deharveng}, {Dixon}, {Kruk},
  {Shull}, {Giroux}, {Morton}, {Williger}, {Friedman}, \& {Moos}}]{Zheng04}
{Zheng}, W. et al., 2004, \apj, 605, 631

\bibitem[{{Zheng} {et~al.}(1997){Zheng}, {Kriss}, {Telfer}, {Grimes}, \&
  {Davidsen}}]{Zheng97}
{Zheng}, W., {Kriss}, G.~A., {Telfer}, R.~C., {Grimes}, J.~P., \& {Davidsen},
  A.~F. 1997, \apj, 475, 469

\end{thebibliography}
\end{document}